\documentclass[prl, twocolumn, showpacs, nofootinbib, superscriptaddress, amsmath,amssymb, floatfix, eqsecnum]{revtex4-1}
\pdfoutput=1
\usepackage{amsmath}
\usepackage{amssymb}
\usepackage{amsthm}
\usepackage{amsfonts}
\usepackage{bbm}
\usepackage{comment}
\usepackage[normalem]{ulem}
\usepackage{graphicx}
\usepackage[dvipsnames]{xcolor}
\usepackage{color,framed}
\usepackage{hyperref}
\usepackage{times}
\usepackage{enumerate}
\usepackage{lipsum}
\usepackage{slashed}
\usepackage{url}
\usepackage{bbm}
\usepackage{chngcntr}
\counterwithout{equation}{section}

\hypersetup{
    colorlinks=true, 
    linktoc=all,     
    linkcolor=blue,  
}

\def \beq {\begin{equation}}
\def \eeq {\end{equation}}
\def \beqa {\begin{eqnarray}}
\def \eeqa {\end{eqnarray}}
\def \bseq {\begin{subequations}}
\def \eseq {\end{subequations}}

\renewcommand{\>}{\rangle}

\begin{document}

\title{Anomaly-free symmetries with obstructions to gauging and onsiteability}
\author{Wilbur Shirley}
\affiliation{Leinweber Institute for Theoretical Physics, University of Chicago, Chicago, Illinois 60637,  USA}
\author{Carolyn Zhang}
\affiliation{Department of Physics, Harvard University, Cambridge, MA 02138, USA}
\author{Wenjie Ji}
\affiliation{Department of Physics and Astronomy, McMaster University, Hamilton, Ontario L8S 4M1, Canada\\
Perimeter Institute for Theoretical Physics, Waterloo, Ontario, N2L 2Y5, Canada}
\author{Michael Levin}
\affiliation{Leinweber Institute for Theoretical Physics, University of Chicago, Chicago, Illinois 60637,  USA}

\begin{abstract}
We present counterexamples to the lore that symmetries that cannot be gauged or made on-site are necessarily anomalous. Specifically, we construct unitary, internal symmetries of two-dimensional lattice models that cannot be consistently coupled to background or dynamical gauge fields or disentangled to a tensor product of on-site operators. These symmetries are nevertheless anomaly-free in the sense that they admit symmetric, gapped Hamiltonians with unique, invertible ground states. We show that symmetries of this kind are characterized by an index $[\omega]\in H^2(G,\mathbb{Q}_+)$, where $\mathbb{Q}_+$ is the multiplicative group of rational numbers labeling one-dimensional quantum cellular automata.
\end{abstract}

\maketitle

\emph{Introduction}---Anomalous symmetries provide powerful constraints on the dynamics of quantum many-body systems. In particular, any Hamiltonian that is invariant under an anomalous symmetry must have nontrivial features at low energies.
In fact, this property can be taken as a \emph{definition} of an anomalous symmetry. 
To be precise, consider a tensor product Hilbert space built out of finite dimensional degrees of freedom arranged in a $d$-dimensional lattice. Assume that the Hilbert space is equipped with a collection of unitary symmetry transformations $\{U_g: g  \in G\}$ parameterized by a finite group $G$ and obeying the group law $U_g U_h \propto U_{gh}$, where $\propto$ indicates equality possibly up to a phase. Assume further that the $\{U_g\}$ transformations are ``quantum cellular automata'' (QCA)\cite{gross2012}: that is, they map strictly local operators to nearby strictly local operators.\footnote{These conditions imply that the $\{U_g\}$ are \emph{internal} symmetries: the finite group condition excludes translational symmetry while the locality preserving condition excludes other crystalline symmetries.} 
In this context, we say that the symmetries $\{U_g\}$ are ``anomaly-free'' if they admit a symmetric, gapped, local Hamiltonian with a unique, invertible\footnote{A state $|\psi\>$ is ``invertible" if there exists another state $|\phi\>$ such that $|\psi\> \otimes |\phi\>$ can be transformed into a product state via a unitary generated by finite time evolution of a local (time-dependent) Hamiltonian.} ground state -- or more generally if they admit such a Hamiltonian after first tensoring with suitable ancilla degrees of freedom.\footnote{Here, the ancillas are required to be finite dimensional and transform under an on-site representation of the symmetry.} 
We say the symmetries are ``anomalous'' otherwise.
Such anomalous symmetries naturally occur in the low energy description of the spatial boundaries of symmetry-protected topological (SPT) phases\cite{callan1985,wen2011,ryu2012electro,chen2013,vishwanath2013,Else:2014vma,Kapustin:2024rrm}.

One class of symmetries that are manifestly anomaly-free are those for which the $U_g$ symmetry transformations are ``on-site", meaning $U_g  = \prod_i U_{g,i}$ where each $U_{g,i}$ is a single-site unitary acting on site $i$ and obeying the group law $U_{g,i} U_{h,i} = U_{gh,i}$. It is easy to see that such on-site symmetries always admit  symmetric, gapped Hamiltonians with product-like ground states, and are therefore anomaly-free according to the above definition.

By a simple extension of this logic, one can argue that any symmetry that is merely ``onsiteable'' is also anomaly-free. Here, we say that a $G$-symmetry is onsiteable if it can be transformed into an on-site symmetry using the following pair of operations: 1) tensoring with an on-site $G$-symmetry acting on an ancillary Hilbert space, and 2) conjugation by a QCA (see Ref.~\onlinecite{zhang2024long} for an example where the first operation is necessary).

Although the above reasoning only works in one direction, it is natural to speculate about the converse statement: if a symmetry is not onsiteable, must it be anomalous? It is not hard to see that the answer is ``yes'' for zero-dimensional systems,\footnote{In the zero-dimensional case, a symmetry is not onsiteable if and only if the $\{U_g\}$ describe a nontrivial projective representation of the group $G$.} and an affirmative answer was also established in the one-dimensional case in Ref.~\onlinecite{seifnashri2025}. In fact, a common piece of lore is that this statement holds generally. 

The purpose of this paper is to present a class of counterexamples to this lore: we construct examples of unitary, finite group symmetries of two-dimensional (2D) lattice systems that are not onsiteable, but are nevertheless anomaly-free. We confirm the absence of anomalies by constructing explicit symmetric, gapped Hamiltonians with unique, gapped invertible ground states. We also show that our examples have the interesting property that they cannot be consistently coupled to a background or dynamical gauge field.\footnote{In fact, this is a closely related property: one can show that an internal symmetry is dynamically gaugeable if and only if it is onsiteable\cite{supp}.}
It is well known that anomalous symmetries present obstructions to coupling to dynamical gauge fields\cite{tHooft:1979rat,Wen:2013oza,Kapustin:2014lwa,kapustin2014,seifnashri2024}, but our examples show that obstructions also occur for some anomaly-\emph{free} symmetries.

\begin{figure}[tbp]
\centering
\includegraphics[width=.7\columnwidth]{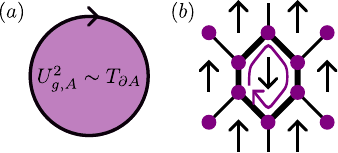} 
\caption{(a) A fundamental property of the $\mathbb{Z}_2$ symmetry (\ref{z2symm}): when restricted to a disk $A$, the restricted symmetry $U_{g,A}$ squares to a translation along $\partial A$ (up to a 1D FDQC along $\partial A$, as indicated by the symbol $\sim$). (b) Action of $\hat{S}$ on vertex qubits.}
\label{fig:dwtransl}
\end{figure}

The class of symmetries we consider share a common feature: when restricted to a finite region, the composition of symmetry transformations becomes ``twisted" by nontrivial translations along the boundary of the region, as illustrated in Fig.~\ref{fig:dwtransl}(a). This phenomenon is captured by an index valued in the cohomology group $H^2(G,\mathbb{Q}_+)$, where $\mathbb{Q}_+$ denotes the multiplicative group of positive rational numbers.\footnote{We note that as a group, $\mathbb{Q}_+$ is isomorphic to a direct sum of infinitely many copies of $\mathbb{Z}$, one for each prime number. That is, $\mathbb{Q}_+\cong\bigoplus_{p\text{ prime}}\mathbb{Z}$.}
We show that a nontrivial index implies an obstruction to both onsiteability and gaugeability.

\emph{A simple example for $G=\mathbb{Z}_2$}---We begin by describing a simple example of a $\mathbb{Z}_2$ symmetry in 2D that is not onsiteable or gaugeable, yet is anomaly-free. Our example is defined on a honeycomb lattice with a qubit on each plaquette and vertex (Fig.~\ref{fig:dwtransl}(b)). We denote the Pauli operators acting on plaquette $p$ by $\sigma_p^{x,y,z}$ and those acting on vertex $i$ by $\tau_i^{x,y,z}$. 

Our $\mathbb{Z}_2$ symmetry operator is defined in terms of a functional $\hat{S}$, which takes as input a configuration of plaquette qubits $\{\sigma^z_p\}$, and outputs a unitary operator that acts on the vertex qubits $\{\tau_i\}$. In particular, for every domain wall $C$ of $\{\sigma^z_p\}$, the unitary $\hat{S}[\{\sigma^z_p\}]$ implements a unit \emph{shift} of all qubits along the loop, $\{\tau_{i}:i\in C\}$. The direction of translation is chosen as follows: along the domain wall and facing the direction of forward translation, the $\uparrow$ domain is always to the left and the $\downarrow$ domain is always to the right (Fig.~\ref{fig:dwtransl}(b)).
With this notation, we define our $\mathbb{Z}_2$ symmetry operator $U_g$ as
\begin{equation}\label{z2symm}
    U_g=\hat{S}[\{\sigma^z_p\}]\prod_p\sigma^x_p.
\end{equation}
The group composition law $U_g^2=\mathbbm{1}$ follows from the fact that $\hat{S}[\{\sigma^z_p\}]=\hat{S}[\{-\sigma^z_p\}]^{-1}$. One can also check that $U_g$ is a QCA: the $\sigma^z_p$ operators transform as $U_g^{-1} \sigma^z_p U_g = - \sigma^z_p$ while the $\sigma_p^{x,y}$ operators transform into operators supported on the six plaquettes and six vertices surrounding $p$. Similarly, $U_g$ maps $\tau_i^{x,y,z}$ into an operator supported on the three plaquettes and four vertices neighboring $i$ (including $i$). We present the explicit transformation laws for each of these operators in the Supplemental Material\cite{supp}. There we also find an explicit representation of $U_g$ as a finite depth quantum circuit (FDQC). 

We now show that $U_g$ (\ref{z2symm}) is anomaly-free. We do this by constructing a $U_g$-symmetric, gapped Hamiltonian that has a product state as its unique gapped ground state.\footnote{Alternatively, one could argue that $U_g$ is anomaly-free simply because there are no nontrivial $\mathbb{Z}_2$ SPTs in 3D\cite{chen2013} and therefore no nontrivial $\mathbb{Z}_2$ anomalies in 2D.} This product state is given by $|\Psi_0\rangle=\otimes_{p,i}|\sigma^x_p=+1,\tau^x_i=+1\rangle$. It is easy to check that $ U_g |\Psi_0\rangle = |\Psi_0\rangle$: indeed, $|\Psi_0\rangle$ is clearly invariant under $\prod_p \sigma^x_p$, and it is also invariant under $\hat{S}$ since the $\tau$ spins are in a fully symmetric state and the $\hat{S}$ operator simply permutes the $\tau$ spins. We claim that the following frustration-free Hamiltonian is gapped and symmetric and has $|\Psi_0\rangle$ as its unique ground state:\footnote{We also construct an alternative \emph{commuting projector} parent Hamiltonian for $|\Psi_0\>$ in \cite{supp}.}
\begin{equation}\label{hamiltonian}
    H=-\sum_i\tau^x_i-\sum_p\sigma^x_p-\sum_pU_g^{-1}\sigma^x_pU_g.
\end{equation}

To see that $[H, U_g] = 0$ notice that $[\sum_i\tau_i^x, U_g] = 0$ while the $\sigma^x_p$ term is explicitly symmetrized. One can also check that $|\Psi_0\>$ is a ground state of $H$: in fact, $|\Psi_0\rangle$ is a ground state of each individual term in $H$.

All that remains is to show that $H$ has a gap to other states. To see this, observe that $[H,\sum_i\tau_i^x] = 0$, so we can diagonalize $H$ separately within each of the different eigenspaces of $\sum_i\tau_i^x$. In fact, we only need to consider the eigenspace that contains $|\Psi_0\rangle$, namely $\sum_i\tau_i^x=N_\mathrm{site}$, since any state in any other eigenspace must have an energy of at least $2$ above $|\Psi_0\rangle$, due to the first term in $H$. At the same time, within the $\sum_i\tau_i^x=N_\mathrm{site}$ eigenspace, the operator $\hat{S}$ acts like the identity since all states in this eigenspace are symmetric under permutations of the $\tau$ spins. Hence $H$ reduces to $H=-N_\mathrm{site}-2\sum_p\sigma^x_p$ within this eigenspace. The claim now follows since the latter expression for $H$ is clearly gapped and has $|\Psi_0\rangle$ as its unique ground state. 

Finally, let us explain the intuitive picture for why $U_g$ (\ref{z2symm}) is not onsiteable. The key idea is to restrict $U_g$ to a disk $A$ and then consider the square of this restriction. 
Here, a ``restriction'' of $U_g$ is a QCA $U_{g,A}$ that acts like $U_g$ deep in the interior of $A$ and acts like the identity far outside of $A$\cite{Else:2014vma}.
In the next section, we find such a restriction of (\ref{z2symm}) and we show that its square, $U_{g,A}^2$, acts like the identity everywhere except near the boundary $\partial A$, where it performs a unit clockwise shift of the $\tau$ spins (Fig.~\ref{fig:dwtransl}(a)). This calculation implies that the square of \emph{any} restriction of (\ref{z2symm}) must implement an \emph{odd} shift of the boundary qubits. Indeed, any other restriction of $U_g$ can be obtained by multiplying $U_{g,A}$ by a one-dimensional (1D) QCA supported near the boundary $\partial A$. Such 1D QCAs can always be written as (integer) clockwise or counterclockwise shifts composed with FDQCs \cite{gross2012}, so they can only contribute an even shift to $U_{g,A}^2$. 
In contrast, for an on-site $\mathbb{Z}_2$ symmetry, there is an obvious way to restrict to a region $A$ such that the restriction squares to the identity. Likewise, the same is true for any onsiteable symmetry. This trivial action is inconsistent with the nontrivial boundary shift action of $U_{g,A}^2$, implying that (\ref{z2symm}) is not onsiteable. 

\emph{The $H^2(G,\mathbb{Q}_+)$ onsiteability index}---We now define an index that formalizes and generalizes the above obstruction to onsiteability. This index is defined for any collection of 2D symmetries $\{U_g\}$ that can be written as FDQCs.
\footnote{In general, FDQC symmetries are only a subset of QCA symmetries, but in 2D systems with finite $G$, it is plausible that the two sets are the same, in view of the theorem of Ref. \onlinecite{Freedman:2019ncn}.} The index takes values in the cohomology group $H^2(G,\mathbb{Q}_+)$ which consists of equivalence classes of functions $\omega : G \times G \to \mathbb{Q}_+$ obeying the cocycle condition
\begin{equation}    \omega(g,h)\omega(gh,k)=\omega(h,k)\omega(g,hk).
\label{omegacocycle}
\end{equation}

Two cocycles $\omega, \omega'$ are considered equivalent if they differ by a coboundary transformation of the form
\begin{equation}
\omega'(g,h)=\alpha(g)\alpha(h)\alpha(gh)^{-1}\omega(g,h),
\label{omegacoboundary}
\end{equation}
where $\alpha$ is a function $\alpha: G \to \mathbb{Q}_+$. 

To define the index, we choose a restriction of each of our symmetry operators $U_g$ to a large disk $A$. These restrictions $\{U_{g,A}\}$ are guaranteed to exist since our symmetries are FDQCs \cite{Else:2014vma}. Next, for every pair of group elements $g, h \in G$ we define the unitary operator
\begin{equation}
    \Omega_{g,h}=U_{g,A}U_{h,A}U_{gh,A}^{-1}.
\end{equation}
By construction, $\Omega_{g,h}$ is supported near $\partial A$, the boundary of $A$. Furthermore, $\Omega_{g,h}$ is a QCA since the $\{U_{g,A}\}$ are QCAs. 

To proceed further, we use the fact that every 1D QCA  $\Omega$ is associated with a positive rational number, known as the ``GNVW index'' of $\Omega$\cite{gross2012}. Roughly speaking, the GNVW index, denoted $\mathrm{Ind}(\Omega)$, is the ratio between the number of degrees of freedom that are shifted clockwise and counterclockwise by $\Omega$ (see \cite{supp} for the precise definition). For example, a unit clockwise or counterclockwise shift on a closed qubit chain has a GNVW index of $2$ or $2^{-1}$ respectively. 

The reason the GNVW index is useful for our purposes is that it completely classifies 1D QCAs: two 1D QCAs have the same GNVW index if and only if they are related by composition with a 1D FDQC (possibly after tensoring with the identity operator on ancillas) \cite{gross2012}. We can therefore use the GNVW index to characterize the unitary operator $\Omega_{g,h}$. Specifically, we define a function $\omega: G \times G \rightarrow \mathbb{Q}_+$ by
\begin{equation}
    \omega(g,h)=\mathrm{Ind}(\Omega_{g,h}).
    \label{index_def}
\end{equation}
It is straightforward to check that $\omega$ obeys the cocycle condition (\ref{omegacocycle}): this follows from associativity of $\{U_{g,A}\}$, which implies the identity \cite{Else:2014vma}
\begin{equation}\label{associativity}
\Omega_{g,h}\Omega_{gh,k}=U_{g,A}\Omega_{h,k}U_{g,A}^{-1}\Omega_{g,hk},
\end{equation}
together with the fact that the GNVW index is multiplicative under composition of unitaries \cite{gross2012}.

At the same time, $\omega$ is only well-defined up to the coboundary transformation (\ref{omegacoboundary}). To see this, recall that there is an ambiguity when choosing a restriction of an FDQC. In particular, if $U_{g,A}$ is a restriction of $U_g$ to region $A$, then we can obtain a different restriction by taking $U_{g,A}' = V_{g, \partial A} U_{g,A}$, where $V_{g, \partial A}$ is a 1D QCA supported near the boundary of $A$. If we now consider the corresponding operator $\Omega'_{g,h}$ and compute its index $\omega'(g,h)$, it is straightforward to check that it differs from $\omega$ by precisely the coboundary transformation (\ref{omegacoboundary}) where $\alpha(g)=\mathrm{Ind}(V_{g, \partial A})$. 
We conclude that the equivalence class $[\omega]\in H^2(G,\mathbb{Q}_+)$ is a well-defined quantity insensitive to the particular choice of restriction.

To see why the index (\ref{index_def}) describes an obstruction to onsiteability, notice that $[\omega]$ is trivial for on-site symmetries: in that case, we can take the natural restriction $U_{g,A} = \otimes_{g\in A}U_{g,i}$ which gives $\Omega_{g,h}=1$ for all $g,h$. Likewise, if $U_{g,A}$ is an on-site symmetry conjugated by a QCA $V$, then we can choose the truncation $U_{g,A} = \otimes_{i\in A}V^\dagger U_{g,i}V$ which also gives $\Omega_{g,h}=1$ for all $g,h$. Therefore, if a symmetry is onsiteable, then $[\omega]\in H^2(G,\mathbb{Q}_+)$ is trivial. Equivalently, if $[\omega]$ is nontrivial, then there is an obstruction to making the symmetry on-site.

\begin{figure}[tbp]
\centering
\includegraphics[width=.8\columnwidth]{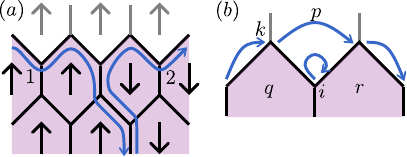}
\caption{(a) Action of $\Omega_{g,g}$ for a typical $\{\sigma_p^z\}$ configuration. Here $A$ is the shaded region and the grey arrows show the upward orientation of $\tilde{\sigma}^z_p$ for $p \notin A$ (\ref{ugrestrict}). The two blue arrows show the
translation action of the two factors of $\hat{S}$ on vertex qubits, with the labels $1,2$ indicating their ordering.
(b) Relationship between plaquettes $q, r$ and vertices $i, k$ in (\ref{eq:OmegaZ2}).}
\label{fig:z2ex}
\end{figure}

To complete the discussion, we now compute our index for the $\mathbb{Z}_2$ symmetry (\ref{z2symm}) discussed earlier. First we choose a restriction $U_{g,A}$ of the symmetry operator $U_g$ to a disk $A$:
\begin{equation}\label{ugrestrict}
    U_{g,A}=\hat{S}[\{\tilde{\sigma}^z_p\}]\prod_{p\in A}\sigma^x_p\qquad\tilde{\sigma}^z_p=\begin{cases}
            \sigma^z_p, \text{ if } p\in A\\
            1, \text{ otherwise}
           \end{cases}.
\end{equation}
This is a valid restriction because $U_{g,A}$ is a QCA that acts like $U_g$ in the interior of region $A$ and acts as the identity operator outside of $A$.
Next, we compute $\Omega_{g,g}$. Using (\ref{ugrestrict}) we have
\begin{equation}
    \Omega_{g,g}=\hat{S}[\{\tilde{\sigma}^z_p\}]\hat{S}[\{\tilde{\bar{\sigma}}^z_p\}]\qquad\tilde{\bar{\sigma}}^z_p=\begin{cases}
            -\sigma^z_p, \text{ if } p\in A\\
            1, \text{ otherwise}
           \end{cases}.
\end{equation}

To understand the operator $\Omega_{g,g}$, it is helpful to consider its action on particular spin configurations. If $\sigma^z_p=1$ for all $p\in A$ then $\hat{S}[\{\tilde{\sigma}^z_p\}]\hat{S}[\{\tilde{\bar{\sigma}}^z_p\}]$ performs a clockwise shift of all qubits along the links on the boundary of $A$ since $\hat{S}[\{\tilde{\sigma}^z_p\}]$ acts trivially while $\hat{S}[\{\tilde{\bar{\sigma}}^z_p\}]$ performs a translation. 
More generally, for any configuration of $\{\sigma^z_p\}$, $\hat{S}[\{\tilde{\sigma}^z_p\}]$ may perform some translations in the bulk of $A$. However, these will be canceled by $\hat{S}[\{\tilde{\sigma}^z_p\}]$ except near the boundary of $A$ [Fig.~\ref{fig:z2ex}(a)]. The end result is that $\Omega_{g,g}$ acts as a unit clockwise shift of the $\tau_i$ qubits along $\partial A$, but with a small modification: the clockwise shift skips over each $\tau_i$ qubit along $\partial A$ that lies between a $\sigma^z = +1$ spin and $\sigma^z= -1$ spin (oriented clockwise). More explicitly, $\Omega_{g,g}$ can be written as
\begin{equation}
    \Omega_{g,g}=T_{\partial A}\prod_{i\in\partial A_-}\big(\textrm{SWAP}_{ik}\big)^{(1+\sigma^z_q)(1-\sigma^z_r)/4},
    \label{eq:OmegaZ2}
\end{equation}
where $T_{\partial A}$ denotes a unit translation of the $\tau_i$ along $\partial A$ and $i\in\partial A_-$ runs over all sites in $\partial A$ whose legs emanate into $A$. The locations of $i,k$ with respect to $q,r$ are indicated in Fig.~\ref{fig:z2ex}(b).

Given that the GNVW index takes the value $2$ for a qubit translation and is invariant under composing with FDQCs like the second term in (\ref{eq:OmegaZ2}), we conclude that $\mathrm{Ind}(\Omega_{g,g})=2\in\mathbb{Q}_+$. At the same time, it is easy to check that the GNVW indices of $\Omega_{1,1}$, $ \Omega_{1,g}$, and $ \Omega_{g,1}$ are all $1$ since all of these operators are simply the identity. We conclude that the $\mathbb{Z}_2$ symmetry (\ref{z2symm}) corresponds to the cocycle $\omega:\mathbb{Z}_2\times\mathbb{Z}_2\to\mathbb{Q}_+$ with values $\omega(1,1)=\omega(1,g)=\omega(g,1)=1$, and $\omega(g,g)=2$. To see that $\omega$ is in fact a nontrivial cocycle, note that the quantity $\omega(g,1)\omega(g,g)$ transforms as $\omega(g,1) \omega(g,g) \to \omega(g,1)\omega(g,g)\alpha(g)^2$ under a shift by a coboundary (\ref{omegacoboundary}). It follows that, for any trivial cocycle, $\omega(g,1)\omega(g,g)$ is the square of a rational number. Since our cocycle has $\omega(g,1)\omega(g,g)=2$, it is nontrivial.

\emph{Obstruction to (background) gauging}---We now show that symmetries with a nonzero $H^2(G,\mathbb{Q}_+)$ index cannot be consistently coupled to a spatial background gauge field. A corollary of this result is that such symmetries also cannot be consistently coupled to a \emph{dynamical} gauge field since the property of ``dynamical gaugeability'' is stronger than ``background gaugeability.'' We explain this point in more detail in the Supplemental Material \cite{supp}.

We begin by giving a precise definition of a ``background gaugeable'' symmetry. As before, we consider the setting of a tensor product Hilbert space defined on a $d$-dimensional lattice $\Lambda$, equipped with a collection of QCA symmetries $\{U_g: g \in G\}$.
We say that a set of symmetries $\{U_g\}$ is ``background gaugeable'' if there exists a collection of QCAs $\{U_{\{g_i\}}\}$ that are defined for any assignment of group elements to lattice sites, $\{g_i: i \in \Lambda\}$ and that satisfy the following properties:
\begin{enumerate}
    \item If $g_i = g$ for all $i$, then $U_{\{g_{i}=g\}} \propto U_g$. \label{prop1}
    \item If $g_i' = g_i h$ for all $i$, then $U_{\{g_i'\}} \propto U_{\{g_i\}} U_h$. \label{prop2}
    \item If $g'_i = g_i$ except for $i \in R$ for some subset $R \subset \Lambda$, then $U_{\{g_i'\}} \propto W U_{\{g_i\}} $ for some unitary $W$ supported within a finite distance of $R$. \label{prop3}
\end{enumerate}
Here, we use `$\propto$' to denote equality up to a phase. This definition deserves a few comments:

1) To motivate this definition, we note that the $U_{\{g_i\}}$ operators provide a canonical way to couple an arbitrary local symmetric Hamiltonian $H$ to a flat\footnote{More generally, we can also couple $H$ to background gauge fields that are not flat, though the prescription is not unique in that case.} spatial background gauge field. Let $\{g_{ij}\}$ be a background gauge field -- that is, an assignment of group elements to links $\<ij\>$ with the identification $g_{ij} = g^{-1}_{ji}$. Suppose that $g_{ij}$ is ``flat'', that is, near each lattice site $k \in \Lambda$ we can find a local assignment of group elements $g_i^{(k)}$, such that $g_{ij} = g_i^{(k)} (g_j^{(k)})^{-1}$ near $k$. Writing the Hamiltonian as $H = \sum_k H_k$, where each $H_k$ is symmetric and supported near $k$, we define the gauged Hamiltonian $H_{\{g_{ij}\}}$ by $H_{\{g_{ij}\}} = \sum_k U_{\{g_i^{(k)}\}}H_k U_{\{g_i^{(k)}\}}^{-1}$. Notice that the ambiguity in the choice of $g_i^{(k)}$'s, namely $g_i^{(k)} \rightarrow g_i^{(k)} h$, drops out in the definition of $H_{\{g_{ij}\}}$ due to property (\ref{prop2}) above. 

2) An important special case of $U_{\{g_i\}}$ is when $g_i = g$ for $i$ in some subset $R \subset \Lambda$ and $g_i = 1$ outside of $R$. In this case we will denote $U_{\{g_i\}}$ by $U_{g,R}$. It follows from properties (\ref{prop1}) and (\ref{prop3}) that $U_{g,R}$ is a \emph{restriction} of $U_g$ to $R$.

We now derive the claimed obstruction to background gaugeability. More precisely, we establish the contrapositive: we show that a background gaugeable 2D symmetry necessarily has a trivial $H^2(G, \mathbb{Q}_+)$ index.

\begin{figure}[tbp]
\centering
\includegraphics[width=.9\columnwidth]{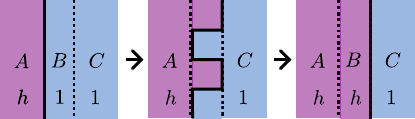} 
\caption{
Graphical representation of $V_1$ (\ref{v1v2def}) as a depth $2$ circuit: the first layer is a product of the $W$ unitaries from property (\ref{prop3}) applied to the purple squares in region $B$. The second layer is a product of the $W$ unitaries acting on the rest of $B$.}
\label{fig:gauging}
\end{figure}

Consider a partition of the lattice $\Lambda$ into three regions $A,B,C$ where $A$ and $C$ are half-planes separated by a large, but finite width strip $B$ (Fig.~\ref{fig:gauging}). For any three group elements $g,h,k \in G$, let $U_{g,h,k}$ denote the operator $U_{\{g_i\}}$ where $g_i$ takes the (constant) values $g$, $h$, and $k$ in the regions $A$, $B$, and $C$, respectively. With this notation, we can state a key identity: 
\begin{align}
U_{g,1,1} U_{h,h,1} \propto  U_{gh,h,1},
\label{key_id}
\end{align}
where `$\propto$' means equality up to phase. To derive this identity, we introduce the notation $U[O] \equiv UOU^{-1}$, and we note that for any operator $O$ supported deep inside $AB$, we have $U_{g,1,1}\circ U_{h,h,1}[O] = U_{g,1,1} \circ U_h[O] = U_{gh,h,1}[O]$ where the second equality follows from properties (\ref{prop2}), (\ref{prop3}) above. Likewise, for any operator $O$ supported deep inside $BC$, we have $U_{g,1,1} \circ U_{h,h,1}[O] = U_{h,h,1}[O] = U_{gh,h,1}[O]$ where the second equality follows from property (\ref{prop3}). Together these two calculations imply the identity (\ref{key_id}).

We now use (\ref{key_id}) to prove that the $H^2(G, \mathbb{Q}_+)$ index is trivial. The idea is that the three operators in (\ref{key_id}) are closely related to the three symmetry restrictions $U_{g, A}$, $U_{h,A}$, and $U_{gh, A}$. Indeed, using property (\ref{prop3}), it is easy to see that 
\begin{align}
U_{h,h,1} = V_1 U_{h,A} , \quad \quad U_{gh,h,1} = V_2 U_{gh,A},
\label{v1v2def}
\end{align}
where $V_1$ and $V_2$ are depth two quantum circuits supported within a finite distance from $B$ (Fig.~\ref{fig:gauging}). Also, we have $U_{g,1,1} = U_{g,A}$, by definition. Therefore all three operators in (\ref{key_id}) are related to the corresponding operators $U_{g, A}$, $U_{h,A}$, and $U_{gh, A}$ by (at most) FDQCs. It then follows from (\ref{key_id}) that $\Omega_{g,h} = U_{g,A} U_{h,A} U_{gh,A}^{-1}$ can be written as an FDQC supported within a finite distance of the quasi-1D strip $B$. Therefore, according to the definition of the index (\ref{index_def}), we have $\omega(g,h) = 1$ for all $g, h$.

\emph{Discussion}---One of the main results of this paper is that every 2D $G$-symmetry that can be written as a finite depth quantum circuit (FDQC) is characterized by an index valued in the cohomology group $H^2(G,\mathbb{Q}_+)$. The reader may wonder about the converse: can all elements of $H^2(G,\mathbb{Q}_+)$ be realized by some symmetry? We answer this question in the affirmative in the Supplemental Material \cite{supp}. There we generalize our $\mathbb{Z}_2$ example (\ref{z2symm}) to construct a 2D anomaly-free $G$-symmetry for \emph{every} cocycle $[\omega]\in H^2(G,\mathbb{Q}_+)$, and \emph{every} group $G$. The construction is based on the mapping between 3D $G$-symmetric QCAs and 2D $G$-symmetries\cite{potter2017,zhang2023}, so along the way we also obtain a new class of $G$-symmetric QCAs in three dimensions. 

Another interesting example of a symmetry with a nontrivial index is the radical Floquet circuit presented in Ref.~\onlinecite{radical} (see also Ref.~\onlinecite{potter2017}). In this 2D Floquet system, the Floquet unitary $U$ is a QCA and obeys $U^4=\mathbbm{1}$, and therefore can be thought of as a $\mathbb{Z}_4$ symmetry transformation. When restricted to a large disk $A$, the restricted symmetry obeys $U_A^4\sim T_{\partial A}^2$ (where $\sim$ means equivalent up to composition with a 1D FDQC), implying that $U$ has a nontrivial index.
In the Supplemental Material \cite{supp}, we study a simplified version of this symmetry $U$, which has the interesting feature that it is an exact (internal) symmetry of the square lattice toric code model that swaps the $e$ and $m$ anyons \cite{Barkeshli:2022wuz}.
We also present several \emph{fermionic} examples in \cite{supp} though we leave a systematic discussion of the fermionic case to future work.

We have emphasized that if a $G$-symmetry has a nontrivial value of the index $[\omega]\in H^2(G,\mathbb{Q}_+)$, then there is an obstruction to making the symmetry on-site or to coupling it to a background (or dynamical) gauge field. It may seem surprising that this obstruction does not correspond to an anomaly in the usual sense. However, there is a key difference: the $H^2(G,\mathbb{Q}_+)$ obstruction is not preserved under renormalization group flow \cite{tHooft:1979rat}. For example, consider a Hilbert space $\mathcal{H}_1 \otimes \mathcal{H}_2$ with symmetry $U_g\otimes U_g^{\mathrm{onsite}}$ where $U_g$ is the $\mathbb{Z}_2$ symmetry defined in (\ref{z2symm}) and $U_g^{\mathrm{onsite}}$ is an on-site symmetry. If we turn on the Hamiltonian $H= H_1 \otimes \mathbbm{1}$ where $H_1$ is given by (\ref{hamiltonian}), then $H$ will gap out all the degrees of freedom in $\mathcal{H}_1$ leaving behind a low energy subspace $\mathcal{H}_2$ with a low energy symmetry $U_g^{\mathrm{onsite}}$. In this example, the original Hilbert space $\mathcal{H}_1 \otimes \mathcal{H}_2$ has an obstruction to onsiteability and gaugeability, while there is no such obstruction in the low energy subspace $\mathcal{H}_2$. 

In this work, we have focused on a particular obstruction to onsiteability that applies to 2D symmetries, but we expect that there are similar obstructions in higher dimensions related to decorating lower dimensional junctions of symmetry defects with nontrivial quantum cellular automata. We also note that we have restricted to Hilbert spaces constructed out of finite local degrees of freedom. Some obstructions including the ones we present here may trivialize in more general Hilbert spaces, for example those constructed out of infinite dimensional rotor degrees of freedom. We leave a more complete classification and analysis of obstructions in such settings to future work.

\emph{Acknowledgments}---We thank Dominic Else, Lukasz Fidkowski, Po-Shen Hsin, Abhinav Prem, Sahand Seifnashri, and Nikita Sopenko for insightful discussions and the audience at the KITP conference ``Generalized Symmetries in Quantum Field Theory" for their feedback. C.Z. is supported by the Harvard Society of Fellows. This research was supported in part by grant NSF PHY-2309135 to the Kavli Institute for Theoretical Physics (KITP), a Simons Investigator grant (W.S and M.L.) by NSERC Discovery Grant Award (RGPIN-2025-07254) and Discovery Launch Supplement Award (DGECR-2025-00059) (W.J.) and by the
Simons Collaboration on Ultra-Quantum Matter (M.L.), which is a grant from the Simons Foundation (651442).

\emph{Note added}---We thank the authors of Ref.~\onlinecite{dominic} for coordinating submission.
The $H^2(G,\mathbb{Q}_+)$ onsiteability index was also noted in \cite{Kapustin:2025nju,Kawagoe:2025ldx}.

\bibliography{onsite_draft}

\end{document}


\title{Supplemental Material \\ ``Anomaly-free symmetries with obstructions to gauging and onsiteability''}

\author{Wilbur Shirley}
\affiliation{Leinweber Institute for Theoretical Physics, University of Chicago, Chicago, Illinois 60637,  USA}
\author{Carolyn Zhang}
\affiliation{Department of Physics, Harvard University, Cambridge, MA 02138, USA}
\author{Wenjie Ji}
\affiliation{Department of Physics and Astronomy, McMaster University, Hamilton, Ontario L8S 4M1, Canada\\
Perimeter Institute for Theoretical Physics, Waterloo, Ontario, N2L 2Y5, Canada}
\author{Michael Levin}
\affiliation{Leinweber Institute for Theoretical Physics, University of Chicago, Chicago, Illinois 60637,  USA}
\maketitle
\tableofcontents

\section{Review of the GNVW index}\label{GNVWreview}
In this section, we review the GNVW index introduced by Gross, Nesme, Vogts, and Werner\cite{gross2012}. This index, denoted $\mathrm{Ind}(U)$, assigns a positive rational number to each one-dimensional quantum cellular automaton (QCA) $U$. Roughly speaking, the rational number $\mathrm{Ind}(U) \in \mathbb{Q}_+$ quantifies how much $U$ pumps local operators in the forward (versus backward) direction along a one dimensional chain. To compute the index, one chooses two adjacent but disjoint intervals $A, B$ that are large with respect to the range of $U$. The GNVW index is then defined as
\begin{equation}
    \mathrm{Ind}(U)=\frac{\eta(U^{-1} \mathcal{A}U,\mathcal{B})}{\eta(\mathcal{A},U^{-1}\mathcal{B}U)}.
\end{equation}
Here $\mathcal{A}, \mathcal{B}$ denote the operator algebras supported in the two intervals $A, B$ while $U^{-1} \mathcal{A} U$ and $U^{-1} \mathcal{B} U$ denote the transformed algebras.  The quantity $\eta(\mathcal{X},\mathcal{Y})$ can be thought of as the ``overlap'' between two operator algebras $\mathcal{X}$ and $\mathcal{Y}$, and is defined as
\begin{equation}
\eta(\mathcal{X},\mathcal{Y})=\sqrt{\sum_{O_x\in\mathcal{X},O_y\in\mathcal{Y}}|\overline{\mathrm{Tr}}(O_x^\dagger O_y)|^2},
\end{equation}
where $\{O_x\}$ and $\{O_y\}$ form a complete orthonormal basis for the algebras $\mathcal{X}$ and $\mathcal{Y}$ respectively: $\overline{\mathrm{Tr}}(O_x^\dagger O_{x'}) = \delta_{xx'}$ and similarly for $O_y$. Also the notation $\overline{\mathrm{Tr}}$ denotes a normalized trace satisfying $\overline{\mathrm{Tr}}(\mathbbm{1})=1$. One can check that $\eta(\mathcal{X},\mathcal{Y})$ is independent of the particular basis that we choose for $\mathcal{X}$ and $\mathcal{Y}$. Also, $\eta(\mathcal{X},\mathcal{Y})$ is bounded below by $\eta(\mathcal{X},\mathcal{Y}) \geq 1$ since the algebras $\mathcal{X}, \mathcal{Y}$ both contain the identity operator. The physical meaning of $\mathrm{Ind}(U)$ is as follows: it counts the square root of the number of operators that get transported from the left of the cut separating $A$ and $B$ to the right of the cut [this contributes to $\eta(U^{-1}\mathcal{A} U,\mathcal{B})$] divided by the square root of the number of operators that get transported across the cut in the opposite direction [this contributes to $\eta(\mathcal{A},U^{-1}\mathcal{B} U)$]. 

It is illuminating to compute the GNVW index for two important classes of QCAs, namely (1) translations and (2) finite depth quantum circuits (FDQCs). In the case of an $n$-fold translation $U$, one can check that $\mathrm{Ind}(U) = d^n$ where $d$ is the dimension of the Hilbert space on each site. On the other hand, in the case of an FDQC $U$, one can show that $\mathrm{Ind}(U) = 1$.

Finally, we should mention an important property of the GNVW index that we use in the main text: this index is \emph{multiplicative} under both composition and tensor products. That is, for any two QCAs $U, V$,
\begin{align}
\mathrm{Ind}(U V) = \mathrm{Ind}(U \otimes V) = \mathrm{Ind}(U) \mathrm{Ind}(V).
\label{mult}
\end{align}

These properties have an important implication. Recall that two QCAs $U, V$ are said to be ``equivalent'' if they are related by composition with a 1D FDQC (possibly after tensoring with the identity operator on ancillas). It follows from the above properties that if $U, V$ are equivalent then $\mathrm{Ind}(U) = \mathrm{Ind}(V)$. In other words, the GNVW index defines an \emph{invariant} on QCAs. In fact, it was shown in Ref.~\onlinecite{gross2012} that the converse also holds: that is, the GNVW index is a \emph{complete} invariant for 1D QCAs.

\section{Additional details related to $\mathbb{Z}_2$ example}
In this section, we provide more details about the $\mathbb{Z}_2$ symmetry $U_g$ discussed in the main text. In particular, we show that $U_g$ can be written as an FDQC and we work out the explicit action of $U_g$ on all local operators. We also show that $U_g$ admits a symmetric commuting projector Hamiltonian with a product state as its unique gapped ground state.

\subsection{Explicit formula for $U_g$ as an FDQC}


We begin by presenting an explicit formula for the $\mathbb{Z}_2$ symmetry $U_g$ as an FDQC. Recall that in the main text, we defined $U_g$ by
\begin{equation}
    U_g=\hat{S}[\{\sigma^z_p\}]\prod_p\sigma^x_p.
\end{equation}
Here $\hat{S}$ is a unitary operator that acts nontrivially on the vertex qubits $\{\tau_i\}$ in a way that depends on the configuration of plaquette qubits $\{\sigma^z_p\}$. Specifically, the operator $\hat{S}[\{\sigma^z_p\}]$ acts on the $\{\tau_i\}$ by performing a unit translation of all the $\tau_i$ qubits that lie along each of the domain walls of the $\{\sigma^z_p\}$. For each domain wall, this translation is in a direction such that the $\uparrow$ domain is to the left and the $\downarrow$ domain is to the right.  

%

\begin{figure}[t]
   \centering
   \includegraphics[width=.45\columnwidth]{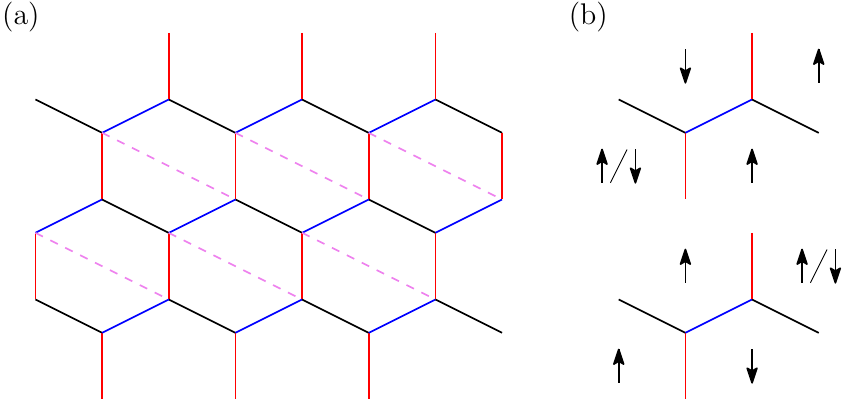}
   \caption{$(a)$ The 2D honeycomb lattice with non-nearest neighbor hopping terms that we use in the FDQC realizing $\hat{S}(\{\sigma^z_p\})$. $(b)$ In the sixth step of the circuit, we apply SWAP gates on every blue edge whose neighboring plaquettes are in one of the spin configurations illustrated here. The plaquettes labeled $\uparrow/\downarrow$ can be in either of the two states $\uparrow$ or $\downarrow$.} 
   \label{fig:translation}
\end{figure}

One might think that any circuit representation of $\hat{S}[\{\sigma^z_p\}]$ needs to have a depth that grows with system size in order to implement the required translations along domain walls. 
However, this is not the case:  $\hat{S}[\{\sigma^z_p\}]$ can be written as a controlled ``SWAP circuit"\cite{po2016} with a finite depth. The key idea is to deform the honeycomb lattice into a square lattice (Fig.~\ref{fig:translation}a) and then perform the SWAP circuit described in Ref.~\onlinecite{po2016}, but with the SWAP gates only applied on certain edges determined by the neighboring
$\{\sigma^z_p\}$ configuration.  

Specifically, the FDQC for $\hat{S}[\{\sigma^z_p\}]$ consists of six steps/layers. The first four steps are a controlled version of the SWAP circuit in Ref.~\onlinecite{po2016}. In the first step, we perform SWAP gates on the red edges shown in Fig.~\ref{fig:translation}a; we then perform SWAP gates on the dotted purple edges in the second step; then SWAP gates on the blue edges in the third step; and finally SWAP gates on the black edges in the fourth step. Each of these SWAP gates are \emph{controlled} gates which are only performed if the neighboring $\sigma^z_p$ spins are in an appropriate configuration. In particular, the SWAP gates on the red, blue, and black edges are performed on all edges except for those that lie between two plaquettes with $\sigma^z_p = \ \downarrow$. Similarly the SWAP gates on the dotted purple edges are performed on all edges except for those that belong to a plaquette with $\sigma^z_p = \ \downarrow$. After these four SWAP circuit-like steps, we need two more steps to complete our circuit representation of $\hat{S}[\{\sigma^z_p\}]$. In the fifth step, we perform SWAP gates on every black edge that is part of a domain wall boundary -- i.e. separates two plaquettes with $\sigma^z_p = \ \uparrow$ and $\sigma^z_p = \ \downarrow$. Finally, in the sixth step we perform SWAP gates on every blue edge whose neighboring plaquettes are in one of the two spin configurations shown in Fig.~\ref{fig:translation}b. 

One can check that the above depth-$6$ circuit exactly realizes the operator $\hat{S}[\{\sigma^z_p\}]$.\footnote{Technically, the gates in each layer described here overlap but mutually commute. If we define an FDQC in terms of layers with gates that have disjoint support, then we must split each of the above six layers into two layers each.} To realize $U_g$, one simply needs to add one more layer to implement the spin flip operation $\prod_p \sigma^x_p$.




\subsection{Action of $U_g$ on local operators}

\begin{figure}[t]
    \centering
    \includegraphics[width=0.12\linewidth]{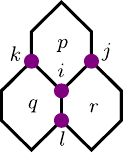}
    \caption{The labeling of the vertices $k,j,l$ and plaquettes $p,q,r$ relative to the site $i$.}
    \label{fig:pqr}
\end{figure}

We now discuss how the Pauli operators $\sigma_p, \tau_i$ transform under $U_g$. 
We begin with the $\sigma_p$ operators. A straightforward calculation shows that
\begin{gather}
    U_g\sigma^z_pU_g^{-1}=-\sigma^z_p,\\
    U_g\sigma^x_pU_g^{-1}=\sigma^x_p
    \Big(T_{\partial p}\prod_{i\in p}\big(\textrm{SWAP}_{ij}\big)^{(\mathbbm{1}+\sigma^z_q)(\mathbbm{1}-\sigma^z_r)/4}\Big)^{\sigma^z_p},\label{eq:sigmax}
\end{gather}
where the sites $i,j$ and plaquettes $q,r$ are defined relative to $p$ as shown in Fig.~\ref{fig:pqr}, and $T_{\partial p}$ is a clockwise shift of the six $\tau$ qubits around $p$.

As for the $\tau_i$ operators, it is easy to see that $\tau_i=\tau^x_i,\tau^y_i,\tau^z_i$ all transform the same way since $U_g$ simply permutes the $\tau$ spins. The explicit transformation law 
can be written as
\begin{equation}
    U_g\tau_iU_g^{-1}=\tau_if_{ii}+\tau_jf_{ij}+\tau_kf_{ik}+\tau_lf_{il},\label{eq:alpha}
\end{equation}
where $j,k,l$ are the sites adjacent to $i$ (see Fig.~\ref{fig:pqr}), and each $f$ is a projector onto certain sets of plaquette spin configurations such that $f_{ii}+f_{ij}+f_{ik}+f_{il}=\mathbbm{1}$. In particular,
\begin{equation}
\begin{gathered}
        f_{ii}=a_pa_qa_r+\bar{a}_p\bar{a}_q\bar{a}_r,\\
    f_{ij}=a_p\bar{a}_r,~~f_{ik}=a_q\bar{a}_p,~~f_{il}=a_r\bar{a}_q,
\end{gathered}
\label{fdef}
\end{equation}
where $p,q,r$ are the plaquettes surrounding $i$, and $a_p = (\mathbbm{1}+\sigma_p^z)/2$ and $\bar{a}_p =(\mathbbm{1}-\sigma_p^z)/2$.

\subsection{Symmetric commuting projector Hamiltonian}\label{commham}
In this section, we construct a symmetric commuting projector Hamiltonian that has
$|\Psi_0\rangle=\otimes_{p,i}|\sigma^x_p=+1,\tau^x_i=+1\rangle$ as its unique gapped ground state. To define the Hamiltonian, it will be helpful to use the projectors $f$ defined in (\ref{fdef}).

We make note of the following properties of the $f$ operators:
\begin{gather}    f_{ij}f_{ik}=\delta_{jk}f_{ij}\label{eq:fdelta},\\
    f_{ij}f_{ji}=0,\label{eq:fijfji}\\
    U_gf_{ij}U_g^{-1}=f_{ji}\label{eq:ftrans}.
\end{gather}
The first property encodes the fact that the qubit on site $i$ cannot flow to multiple locations. The second says that qubits cannot simultaneously flow from $i$ to $j$ and $j$ to $i$. The third property reflects the fact that a global spin flip reverses the direction of qubit flow. We then define the commuting projector Hamiltonian as
\begin{gather}
    H=-\sum_p \frac{1}{2}(\mathbbm{1} + \sigma^x_p) P_p, \label{Hcommdef}\\
    P_p=\prod_{i\in p}P_{ii}\cdot\prod_{\langle ij\rangle\in p}P_{ij}\cdot\prod_{\langle ij\rangle\perp p}P_{ij},
    \label{Ppdef}
\end{gather}
where $\langle ij\rangle\perp p$ indexes the six links emanating from $p$, and
\begin{equation}\label{pijdef}
    P_{ij}=\begin{cases}
			\textstyle\frac{1}{2}(\mathbbm{1}+\tau^x_j)f_{ij}+\frac{1}{2}(\mathbbm{1}+\tau^x_i)f_{ji}+(\mathbbm{1}-f_{ij}-f_{ji}) & \text{if $i\neq j$}\\
            \frac{1}{2}(\mathbbm{1}+\tau^x_i)f_{ii}+(\mathbbm{1}-f_{ii}) & \text{if $i=j$.}
		 \end{cases}
\end{equation} 
Our main task is to show that the $\frac{1}{2}(\mathbbm{1}+\sigma^x_p) P_p$ operators in (\ref{Hcommdef}) are $U_g$-symmetric, Hermitian commuting projectors. To this end, we note that the $P_{ij}$ operators obey the following properties:
\begin{enumerate}[(i)]
\item{$P_{ij}=P_{ji}$.} \label{proppij1}
\item{$[P_{ij},P_{kl}]=0$ for any $i,j,k,l$.} \label{proppij2}
\item{$P_{ij}$ is a projector.} \label{proppij3}
\item{$ U_g P_{ij} U_g^{-1} = P_{ij}.$} \label{proppij4}
\item{$\tau^x_iP_{ii}P_{ij}P_{ik}P_{il}=P_{ii}P_{ij}P_{ik}P_{il}$ for any $j,k,l$ neighboring $i$.} \label{proppij5}
\end{enumerate}
Here, property (\ref{proppij3}) follows from the fact that $P_{ij}$ is a sum of mutually commuting projectors whose pairwise products are zero, while property (\ref{proppij4}) follows from the transformation laws (\ref{eq:alpha}) and (\ref{eq:ftrans}) combined with (\ref{eq:fdelta}). To derive property (\ref{proppij5}), note that the following identity holds for any neighboring $i, j$: $f_{ji}(\mathbbm{1}-\tau^x_i) P_{ij} =0$ (which follows from (\ref{eq:fdelta})-(\ref{eq:fijfji})). From this identity, together with the related identity $f_{ii} (\mathbbm{1}- \tau^x_i) P_{ii} = 0$, we deduce that 
\begin{displaymath}
(f_{ii}+f_{ji} + f_{ki}+f_{li})(\mathbbm{1}-\tau^x_i)P_{ii}P_{ij}P_{ik}P_{il} = 0,
\end{displaymath}
which implies property (\ref{proppij5}) since $f_{ii}+f_{ji} + f_{ki}+f_{li} = \mathbbm{1}$.

Next using property (\ref{proppij5}) we rewrite $P_p$ (\ref{Ppdef}) in an alternative form that will be useful below:
\begin{align}
P_p &= \prod_{i\in p}\textstyle\frac{1}{2}(\mathbbm{1}+\tau^x_i) \cdot P_p \nonumber \\
&= \prod_{i\in p}\textstyle\frac{1}{2}(\mathbbm{1}+\tau^x_i)\cdot\displaystyle\prod_{\langle ij\rangle\perp p}P_{ij}.
\label{simpp}
\end{align}
Here the first line follows from property (\ref{proppij5}) which guarantees that $\tau_i^x P_p = P_p$ for all $i \in p$. The second line follows from the identities $(\mathbbm{1}+\tau_i^x)P_{ii} = (\mathbbm{1}+\tau_i^x)$ and $(\mathbbm{1}+\tau_i^x)(\mathbbm{1}+\tau_j^x)P_{ij} = (\mathbbm{1}+\tau_i^x)(\mathbbm{1}+\tau_j^x)$ which allows us to remove the factors of $\prod_{i \in p} P_{ii}$ and $\prod_{\langle ij\rangle \in p} P_{ij}$ in the definition of $P_p$ (\ref{Ppdef}).

Similarly, we can derive a simplified expression for the product $P_p P_q$ for any two neighboring plaquettes $p,q$:
\begin{equation}\label{simppq}
    P_pP_q=\prod_{i\in p\cup q}\textstyle\frac{1}{2}(\mathbbm{1}+\tau^x_i)\cdot\displaystyle\prod_{\langle ij\rangle\perp p\cup q}P_{ij}.
\end{equation}

We now have all the results we need to show that the $\frac{1}{2}(\mathbbm{1}+\sigma^x_p) P_p$ operators are Hermitian, commuting projectors. First, note that the expression for $P_p$ given in (\ref{simpp}) implies that $[\sigma^x_p, P_p] = 0$ since the right hand side of (\ref{simpp}) only includes the six $\sigma^z_p$ operators adjacent to $p$. It then follows that $\frac{1}{2}(\mathbbm{1}+\sigma^x_p) P_p$ is a Hermitian projector, since $\frac{1}{2}(\mathbbm{1}+\sigma^x_p)$ and $P_p$ commute with one another and are both Hermitian projectors. Similarly, note that (\ref{simppq}) implies that $[\sigma_p^x, P_p P_q] = 0$ for any two neighboring plaquettes $p, q$. It then follows that $\sigma^x_pP_p$ and $\sigma^x_qP_q$ commute for any neighboring $p, q$:
\begin{equation}\label{termscomm}
    (\sigma^x_pP_p)(\sigma^x_qP_q) = P_pP_q \sigma^x_p \sigma^x_q = P_qP_p \sigma^x_q \sigma^x_p =
    (\sigma^x_qP_q)(\sigma^x_pP_p).
\end{equation}
At the same time, it is also clear that $\sigma^x_pP_p$ and $\sigma^x_qP_q$ commute if $p, q$ are \emph{not} adjacent, so we conclude that the $\frac{1}{2}(\mathbbm{1}+\sigma^x_p) P_p$ terms are indeed Hermitian commuting projectors.

To see that the $\frac{1}{2}(\mathbbm{1}+\sigma^x_p) P_p$ operators are symmetric under $U_g$ note that
\begin{align}
    U_g (\sigma^x_p P_p) U_g^{-1} &= U_g \sigma^x_p U_g^{-1} P_p \nonumber  \\
    &= \sigma^x_p
    \Big(T_{\partial p}\prod_{i\in p}\big(\textrm{SWAP}_{ij}\big)^{(\mathbbm{1}+\sigma^z_q)(\mathbbm{1}-\sigma^z_r)/4}\Big)^{\sigma^z_p} P_p \nonumber \\
    &= \sigma^x_p P_p.
\end{align}
Here, the first line follows from $[U_g, P_p] = 0$ which in turn follows from property (\ref{proppij4}) and (\ref{Ppdef}). The last line follows from the expression for $P_p$ given in (\ref{simpp}): all the SWAP and translation operators act trivially when applied to $P_p$ since this term projects into the $\tau^x_i = 1$ subspace which is symmetric under permutations of $\tau$ spins along the boundary of $p$. 

All that remains is to show that $H$ has $|\Psi_0\rangle=\otimes_{p,i}|\sigma^x_p=+1,\tau^x_i=+1\rangle$ as its unique gapped ground state. To see that $|\Psi_0\rangle$ is a ground state, note that $\sigma^x_p |\Psi_0\rangle = P_p |\Psi_0\rangle = |\Psi_0\rangle$ so $|\Psi_0\rangle$ minimizes every term in the Hamiltonian. To see that it is unique, note that any state with the same energy must also obey $\sigma^x_p = P_p = 1$ and therefore also $\tau_i^x = 1$, in view of (\ref{simpp}).

\section{Dynamical gauging and onsiteability}
In this section, we define the notion of a ``dynamically gaugeable'' symmetry, and we show that every dynamically gaugeable symmetry is also background gaugeable according to the definition given in the main text. One implication of this result is that the $H^2(G, \mathbb{Q}_+)$ obstruction to background gaugeability is also an obstruction to dynamical gaugeability. 

We also derive a direct connection between dynamical gaugeability and onsiteability: we show that a finite, internal symmetry is dynamically gaugeable if and only if it is onsiteable. This result means that quite generally, any such symmetry that is not onsiteable has an obstruction to dynamical gauging, whether or not it is anomalous.


\subsection{Definition of dynamical gaugeability}\label{sgausslaw}

We first define what we mean for a symmetry to be dynamically gaugeable. Consider a tensor product Hilbert space $\mathcal{H}_\text{matter}=\bigotimes_{i\in\Lambda}\mathcal{H}_i$ defined on a $d$-dimensional lattice $\Lambda$, and equipped with a set of QCA symmetries $\{U_g : g \in G\}$. Define an expanded Hilbert space $\mathcal{H}=\mathcal{H}_\text{matter}\otimes\mathcal{H}_\text{gauge}$, where $\mathcal{H}_\text{gauge}=\bigotimes_{\langle ij\rangle}\mathbb{C}^{|G|}$ consists of a $G$-qudit on each link of $\Lambda$. As in standard lattice gauge theory\cite{kogut1979}, we think of $\mathcal{H}_{\text{gauge}}$ as describing the gauge field degrees of freedom and we denote the basis states of $\mathcal{H}_\text{gauge}$ by $\ket{\{g_{ij}\}}$, where $g_{ji}=\bar{g}_{ij}$ (here, we use the notation $\bar{g}\equiv g^{-1}$). For each qudit $\langle ij\rangle$, we let $P_{g,ij}$ denote the projector onto the state $\ket{g}$, i.e.~$P_{g}\ket{h}=\delta_{gh}\ket{g}$, and we let $L_{g,ij}$ denote the left multiplication operator, $L_{g}|h\rangle = |gh\rangle$, and $R_{g,ij}$ denote the right multiplication operator, $R_{g}|h\rangle = |hg\rangle$ (note that $L_{g,ij} = R_{\bar{g},ji}$). Finally, we let $\mathsf{L}_{g,i}=\prod_{\langle ij\rangle}L_{g,ij}$ denote the product of $L_{g,ij}$ over all links emanating from site $i$. 

We say that the symmetry $\{U_g\}_{g\in G}$ is dynamically gaugeable if there exists a collection of unitary ``Gauss's law'' operators $\{\Gs_{g,i}\}_{g\in G,i\in\Lambda}$ acting on $\mathcal{H}$, that satisfy the following properties:
\begin{enumerate}[\qquad G1.]
    \item $\Gs_{g,i}=\mathsf{L}_{g,i} \cdot \mathsf{V}_{g,i}$ where $\mathsf{V}_{g,i}$ is a unitary operator that is diagonal in the $|\{g_{ij}\}\rangle$ basis and supported in a finite disk centered at $i$. \label{G1}
    \item $\Gs_{g,i}\Gs_{h,i}=\Gs_{gh,i}$. \label{G2}
    \item $\Gs_{g,i}\Gs_{h,j} = \Gs_{h,j}\Gs_{g,i}$ for $i\neq j$. \label{G3}
    \item $\bra{\{g_{ij} = 1\}}\prod_i\Gs_{g,i}\ket{\{g_{ij} = 1\}}\propto U_g$. \label{G4}
\end{enumerate}

The motivation for requiring these properties is as follows. Property (G1) ensures that the Gauss's law operators are local and transform the gauge fields in the usual way: $\Gs_{g,i}P_{h,ij}(\Gs_{g,i})^{-1}=P_{gh,ij}$ and $\Gs_{g,i}P_{h,jk}\Gs_{g,i}^{-1}=P_{h,jk}$ for $i\neq j,k$. Properties (G2) and (G3)  guarantee that the sum over all gauge transformations, $\frac{1}{|G|^\Lambda} \sum_{\{g_i\}} \prod_i G_{g_i,i}$, is a projection operator onto the gauge invariant subspace where $\mathsf{G}_{g,i} = 1$, just as in standard lattice gauge theory.
Finally, property (G4) ensures that the Gauss's law operators are related to the global symmetry operators in the usual way. Here, in (G4) we are using a notation where the expression
$\bra{\{g_{ij} = 1\}} \prod_i\Gs_{g,i} \ket{\{g_{ij} = 1\}}$ denotes the unique operator acting on $\mathcal{H}_\text{matter}$ whose matrix element between two states $\ket{\Psi}, \ket{\Psi'} \in \mathcal{H}_\text{matter}$ is given by
\begin{align}
 \bra{\Psi'}(\bra{\{g_{ij} = 1\}} \textstyle \prod_i\Gs_{g,i} \ket{\{g_{ij} = 1\}})\ket{\Psi} =   (\bra{\Psi'}\otimes \bra{\{g_{ij} = 1\}}) \textstyle \prod_i\Gs_{g,i} (\ket{\{g_{ij} = 1\}} \otimes \ket{\Psi}) .
\end{align}
As in the main text, the `$\propto$' symbol in (G4) denotes equality up to a phase. 

Given any collection of Gauss's law operators $\{\Gs_{g,i}\}$ obeying the above properties, one can construct a dynamical gauge theory in the same way as in standard lattice gauge theory\cite{kogut1979}: first, one defines the Hilbert space for the dynamical gauge theory to be the subspace of $\mathcal{H}$ where $\Gs_{g,i} = 1$ for all $g, i$ (i.e.~the ``gauge invariant subspace''). Then, one chooses a Hamiltonian $H$ that commutes with the $\{\Gs_{g,i}\}$ (i.e.~a ``gauge invariant'' Hamiltonian). Finally, one studies the spectrum of $H$ within the $\Gs_{g,i} = 1$ subspace.

The simplest symmetries that satisfy the above requirements are \emph{on-site} symmetries: for any on-site symmetry, $U_g = \prod_i U_{g,i}$, one can construct suitable Gauss's law operators by choosing $\mathsf{V}_{g,i}$ to be the single site symmetry transformation, $\mathsf{V}_{g,i}=U_{g,i}$\cite{kogut1979}. By construction, these Gauss's law operators satisfy all of the above properties, (G1-G4). More generally, the same argument implies that every \emph{onsiteable} symmetry is also dynamically gaugeable: one can always put such symmetries into an on-site form (at least after adding suitable ancillas) and then use this on-site representation to construct Gauss's law operators as above. 

\subsection{Proof that dynamical gaugebility implies background gaugeability}
Recall that in the main text we defined a set of QCA symmetries $\{U_g : g \in G\}$ to be ``background gaugeable'' if there exists a set of QCAs $\{U_{\{g_i\}}\}$, acting on $\mathcal{H}_{\mathrm{matter}}$, obeying the following properties: 
\begin{enumerate}
    \item If $g_i = g$ for all $i$, then $U_{\{g_{i}=g\}} \propto U_g$. \label{prop1}
    \item If $g_i' = g_i h$ for all $i$, then $U_{\{g_i'\}} \propto U_{\{g_i\}} U_h$. \label{prop2}
    \item If $g'_i = g_i$ except for $i \in R$ for some subset $R \subset \Lambda$, then $U_{\{g_i'\}} \propto W U_{\{g_i\}} $ for some unitary $W$ supported within a finite distance of $R$. \label{prop3}
\end{enumerate}
We now show that dynamical gaugeability implies background gaugeability. That is, we show that if there exists a set of Gauss law operators $\{\Gs_{g,i}\}$ satisfying (G1-G4), then there exists a set of QCAs $\{U_{\{g_i\}}\}$ acting on $\mathcal{H}_{\mathrm{matter}}$ satisfying the above properties (\ref{prop1})-(\ref{prop3}). 

Our goal is to construct $\{U_{\{g_i\}}\}$ given $\{\Gs_{g,i}\}$. As a first step, we define operators $Y_{\{g_i\}}$, acting on $\mathcal{H}_{\text{matter}}\otimes\mathcal{H}_{\text{gauge}}$, that perform gauge transformations. These operators are defined as the product of Gauss's law operators:
\begin{align}
Y_{\{g_i\}}=\prod_i\Gs_{g_i,i}.
\label{ydef}
\end{align}
Note that ordering in the above product does not matter due to (G3). Next we define $U_{\{g_i\}}$ by considering the action of $Y_{\{g_i\}}$ on the $g_{ij} = 1$ subspace:
\begin{equation}
U_{\{g_i\}} \equiv \bra{\{g_{ij}=g_i\bar{g}_j\}} Y_{\{g_i\}} \ket{ \{g_{ij} =1\}}.
\label{udef}
\end{equation}

It is clear from the above definition that $U_{\{g_i\}}$ is unitary. To show that it is a QCA, we will derive an explicit representation of $U_{\{g_i\}}$ as a depth-$n$ quantum circuit. The first step is to partition the lattice $\Lambda$ into $n$ sublattices $\Lambda_1,\cdots \Lambda_n$ such that the $\Gs_{g_i,i}$ operators in each sublattice have non-overlapping regions of support: that is, if $i,j\in \Lambda_m$ then $\mathrm{supp}(\Gs_{g,i})\cap \mathrm{supp}(\Gs_{h,j})=\emptyset$.\footnote{Here, $n$ scales like the number of lattice sites contained in the region of support of the $\Gs_{g,i}$ operators. In particular, $n$ is finite.} We then use this partition to rewrite $U_{\{g_i\}}$ as
\begin{align}
U_{\{g_i\}} & =\bra{\{g_{ij}=g_i\bar{g}_j\}}   \textstyle \left(\prod_{i\in\Lambda_1}\Gs_{g_i,i}\right)\cdots \left(\prod_{i\in\Lambda_n}\Gs_{g_i,i}\right) \ket{ \{g_{ij} =1\}} \nonumber \\
& =\bra{\{g_{ij}=g_i\bar{g}_j\}}   \textstyle \left(\prod_{i\in\Lambda_1}\mathsf{L}_{g_i,i} \mathsf{V}_{g_i,i}\right)\cdots \left(\prod_{i\in\Lambda_n}\mathsf{L}_{g_i,i} \mathsf{V}_{g_i,i}\right) \ket{ \{g_{ij} =1\}} \nonumber \\
&= \bra{\{g_{ij}=g_i\bar{g}_j\}} \textstyle ( \prod_i\mathsf{L}_{g_i,i}) (\prod_{i\in\Lambda_1}\tilde{\mathsf{V}}_{\{g_j\},i})\cdots (\prod_{i\in\Lambda_n}\tilde{\mathsf{V}}_{\{g_j\},i}) \ket{ \{g_{ij} =1\}} \nonumber \\
&= \bra{\{g_{ij}=1\}} \textstyle (\prod_{i\in\Lambda_1}\tilde{\mathsf{V}}_{\{g_j\},i})\cdots (\prod_{i\in\Lambda_n}\tilde{\mathsf{V}}_{\{g_j\},i}) \ket{ \{g_{ij} =1\}},
\label{Ugider}
\end{align}
where $\tilde{\mathsf{V}}_{\{g_j\},i}$ is defined by 
\begin{equation}
    \tilde{\mathsf{V}}_{\{g_j\},i}=\left(\prod_{j\in I_{m+1}\cup\cdots\cup I_n}\mathsf{L}_{g_j,j}\right)^\dagger\mathsf{V}_{g_i,i}\left(\prod_{j\in I_{m+1}\cup\cdots\cup I_n}\mathsf{L}_{g_j,j}\right),
\end{equation}
for $i\in I_m, m<n$ and $\tilde{\mathsf{V}}_{\{g_j\},i}=\mathsf{V}_{\{g_j\},i}$ for $i\in I_n$. Next, we note that the $\tilde{\mathsf{V}}_{\{g_j\},i}$ operators are diagonal in the $|\{g_{ij}\} \rangle$ basis (just like the $\mathsf{V}_{g_i,i}$ operators) so we can rewrite (\ref{Ugider}) as
\begin{align}
U_{\{g_i\}} =  \left(\prod_{i\in\Lambda_1} Z_{\{g_j\},i} \right)\cdots \left(\prod_{i\in\Lambda_n}Z_{\{g_j\},i} \right),
\label{Ugider2}
\end{align}
where $Z_{\{g_j\},i}$ is a unitary operator acting on $\mathcal{H}_{\mathrm{matter}}$, defined by
\begin{align}
Z_{\{g_j\},i} \equiv \bra{\{g_{ij}=1\}} \textstyle \tilde{\mathsf{V}}_{\{g_j\},i} \ket{ \{g_{ij} =1\}}.
\end{align}
Finally, notice that the region of support of $Z_{\{g_j\},i}$ is contained within that of $\Gs_{g_i,i}$ so in particular, the $Z_{\{g_j\},i}$ are local unitary operators with non-overlapping regions of support within each sublattice $\Lambda_m$. It follows that Eq.~(\ref{Ugider2}) is an explicit realization of $U_{\{g_i\}}$ as a depth-$n$ quantum circuit.

Let us now check that the operators $\{U_{\{g_i\}}\}$ defined by (\ref{udef}) satisfy conditions (\ref{prop1}-\ref{prop3}). First, observe that
\begin{equation}
\bra{\{g_{ij} = 1\}} Y_{\{g_i=g\}} \ket{\{g_{ij} = 1\}} \propto U_g,
\end{equation}
according to (G4), so condition (\ref{prop1}) is satisfied. Also, using (G2-G3), we find that 
\begin{equation}\label{ycommute}
\bra{\{g_{ij} = g_i \bar{g}_j\}} Y_{\{g_ih\}} \ket{\{g_{ij} = 1\}} = \bra{\{g_{ij} = g_i \bar{g}_j\}} Y_{\{g_i\}}Y_{\{g_i=h\}} \ket{\{g_{ij} = 1\}}.
\end{equation}
From (\ref{ycommute}) we deduce that condition (\ref{prop2}) is satisfied: 
\begin{align}
    U_{\{g_ih\}} &= \bra{\{g_{ij} = g_i \bar{g}_j\}} Y_{\{g_ih\}}\ket{\{g_{ij} = 1\}} \nonumber \\
    &=\bra{\{g_{ij} = g_i \bar{g}_j\}} Y_{\{g_i\}}Y_{\{g_i=h\}} \ket{\{g_{ij} = 1\}} \nonumber \\
    &= \bra{\{g_{ij} = g_i \bar{g}_j\}} Y_{\{g_i\}} \ket{\{g_{ij} = 1\}} \bra{\{g_{ij} = 1\}}Y_{\{g_i=h\}} \ket{\{g_{ij} = 1\}} \nonumber \\
    & \propto U_{\{g_i\}}U_h.
\end{align} 

As for condition (\ref{prop3}), suppose that $g_i' = g_i$ except for $i$ in some subset $R \subset \Lambda$. We wish to construct a unitary $W$ supported within a finite distance of $R$ such that $U_{\{g_i'\}} = U_{\{g_i\}} W$. To do this, we note that (G2-G3) implies that
\begin{align}
U_{\{g_i'\}} &=
    \bra{\{g_{ij} = g_i' \bar{g}_j'\}}Y_{\{g_i'\}}\ket{\{g_{ij} = 1\}} \nonumber \\
    &= \bra{\{g_{ij} = g_i' \bar{g}_j'\}}Y_{\{\bar{g}_ig_i'\}} Y_{\{g_i\}}\ket{\{g_{ij} = 1\}} \nonumber \\
    &= \bra{\{g_{ij} = g_i' \bar{g}_j'\}}Y_{\{\bar{g}_ig_i'\}} \ket{\{g_{ij} = g_i \bar{g}_j\}}\bra{\{g_{ij} = g_i \bar{g}_j\}}Y_{\{g_i\}}\ket{\{g_{ij} = 1\}}.
\end{align}
It follows that
\begin{align}
    U_{\{g_i'\}} = W U_{\{g_i\}},
\end{align}
where $W$ is a unitary acting on $\mathcal{H}_{\mathrm{matter}}$ defined by 
\begin{align}
W \equiv \bra{\{g_{ij} = g_i' \bar{g}_j'\}}Y_{\{\bar{g}_ig_i'\}} \ket{\{g_{ij} = g_i \bar{g}_j\}}.    
\end{align}
Furthermore, since $Y_{\{\bar{g}_ig_i'\}}$ is supported within a finite distance of $R$, it is clear that $W$ also has this property.

\subsection{Equivalence between dynamical gaugeability and onsiteability}

In this section, we show that a collection of QCA symmetries $\{U_g : g \in G\}$ is dynamically gaugeable if and only if it is onsiteable. One direction is obvious: as we explained in Sec.~\ref{sgausslaw}, it is clear that every onsiteable symmetry is dynamically gaugeable. Our main task is to prove the converse -- that is, every dynamically gaugeable symmetry is onsiteable. We accomplish this using a construction from Ref.~\onlinecite{seifnashri2025}. 


\subsubsection{Modified Gauss's law operators}

We will establish onsiteability by constructing an explicit ``symmetry disentangler'' -- a QCA that maps $\{U_g\}$ to an on-site symmetry, via conjugation, after tensoring with an on-site representation of the symmetry on ancillas. This disentangler acts in an enlarged Hilbert space $\mathcal{H}' \supset \mathcal{H}_\text{matter}$, which is similar to, but distinct from the dynamical gauging Hilbert space $\mathcal{H}$ discussed in Sec.~\ref{sgausslaw}. In particular, while in Sec.~\ref{sgausslaw} we added $G$-qudits on the links of $\Lambda$, now we add $G$-qudits on every \emph{site}. That is, we define $\mathcal{H}'=\mathcal{H}_\text{matter}\otimes\mathcal{H}_\text{ancilla}$, where $\mathcal{H}_\text{ancilla}=\bigotimes_{i\in\Lambda}\mathbb{C}^{|G|}$ is composed of a $G$-qudit on each site of $\Lambda$. We denote the basis states of $\mathcal{H}_\text{ancilla}$ by $\ket{\{s_i\}}$ ($s\in G$). Note that $\mathcal{H}_{\text{ancilla}}$ comes equipped with a natural on-site $G$ symmetry action $\mathcal{L}_g=\prod_i \mathcal{L}_{g,i}$ that acts by left multiplication: $\mathcal{L}_{g}|\{s_i\}\rangle = |\{gs_i\}\rangle$.

The key tool for building our disentangler is a set of ``modified Gauss's law operators'' $\{\G_{g,i}\}_{g\in G,i\in\Lambda}$ acting on $\mathcal{H}'$. We define these modified operators by mapping the original Gauss's law operators $\{\Gs_{g,i}\}$ operators into $\mathcal{H}'$ in a natural way. Specifically, we define
\begin{equation}    
\bra{\{s'_j\}}\G_{g,i} \ket{\{s_j\}} \equiv \begin{cases}
\bra{\{\bar{s}'_js'_k\}} \Gs_{g,i}\ket{\{\bar{s}_js_k\}}, & \text{if $s_i' = s_i \bar{g}$ and $s_j' = s_j$ for $j \neq i$} \\
0, & \text{otherwise}
\end{cases}
\label{gdef}
\end{equation}
By construction, the $\{\G_{g,i}\}$ are symmetric under $\mathcal{L}_g$: 
\begin{align}
     [\mathcal{L}_g,\G_{h,i}] =0.
     \label{Gsymmprop}
\end{align}
Furthermore, one can check that the $\{\G_{g,i}\}$ obey properties analogous to (G1-G4):
\begin{enumerate}[\qquad M1.]
    \item $\G_{g,i} = \mathcal{R}_{\bar{g},i} \cdot \mathcal{V}_{g,i}$, where $\mathcal{R}_{\bar{g},i}$ is the single qudit operator that acts by right multiplication by $\bar{g}$ on site $i$, and $\mathcal{V}_{g,i}$ is a unitary operator that is diagonal in the $|\{s_i\}\rangle$ basis and supported in a finite disk centered at $i$.
    \item $\G_{g,i}\G_{h,i}=\G_{gh,i}$.
    \item $\G_{g,i} \G_{h,j}= \G_{h,j} \G_{g,i}$ for $i\neq j$.
    \item $\bra{\{s_i = 1\}} \mathcal{L}_g \prod_i\G_{g,i}\ket{\{s_i = 1\}}\propto U_g$.
\end{enumerate}
Here, property (M1) follows immediately from (G1), while (M2-M4) follow from (G2-G4) together with the identity
\begin{align}
\bra{\{s'_j\}}\G_{g_1,i_1} \cdots \G_{g_n,i_n} \ket{\{s_j\}} = \begin{cases}
\bra{\{\bar{s}'_js'_k\}} \Gs_{g_1,i_1} \cdots \Gs_{g_n,i_n}\ket{\{\bar{s}_js_k\}}, & \text{if $s_{i_k}' = s_{i_k} \bar{g}_k$ and $s_j' = s_j$ for $j \neq i_1,...,i_n$} \\
0, & \text{otherwise}
\end{cases}
    \label{phiGprop}
\end{align}

\subsubsection{Symmetry disentangler}

We are now ready to construct our symmetry disentangler. More specifically, we will construct an FDQC $\W$, acting on $\Hc'$, that transforms $U_g \otimes \mathcal{L}_g$ into the on-site symmetry $\mathbbm{1} \otimes \mathcal{L}_g$:
\begin{equation}    \W^\dagger(U_g\otimes\mathcal{L}_g)\W\propto\mathbbm{1}\otimes\mathcal{L}_g.
    \label{eq:trans}
\end{equation}
We define $\W$ by specifying its matrix elements:
\begin{equation}    
\bra{\{s_j'\}}\W\ket{\{s_j\}}\equiv  \bra{\{1\}}\prod_i\G_{s_i,i}\ket{\{s_j\}} \delta_{\{s_j'\}, \{s_j\}},
\label{Wdef}
\end{equation}
where $\delta_{\{s_j'\}, \{s_j\}}$ is a Kronecker delta symbol that evaluates to $1$ if $s_j' = s_j$ for all $j$ and is $0$ otherwise. Note that $\W$ is diagonal in the $\ket{\{s_j\}}$ basis. 

To show that $\W$ defined in this way satisfies (\ref{eq:trans}), we compute the nonzero matrix elements of the expression on the left:
\begin{align}
    \bra{\{gs_j\}}\W^\dagger(U_g\otimes\mathcal{L}_g)\W\ket{\{s_j\}}&=
    \bra{\{gs_j\}}\W^\dagger \ket{\{gs_j\}} \bra{\{gs_j\}} (U_g\otimes\mathcal{L}_g) \ket{\{s_j\}} \bra{\{s_j\}} \W\ket{\{s_j\}} \nonumber \\
    &\propto \bra{\{gs_j\}}(\textstyle\prod_i\G_{gs_i,i})^\dagger\ket{\{1\}}\bra{\{1\}} \mathcal{L}_g \prod_i\G_{g,i}\ket{\{1\}}\bra{\{1\}}\prod_i\G_{s_i,i}\ket{\{s_j\}} \nonumber \\
    &= \bra{\{gs_j\}}(\textstyle\prod_i\G_{gs_i,i})^\dagger \mathcal{L}_g \prod_i\G_{g,i}\prod_i\G_{s_i,i}\ket{\{s_j\}} \nonumber \\
    &=\bra{\{s_j\}}(\textstyle\prod_i\G_{gs_i,i})^\dagger\prod_i\G_{g,i}\prod_i\G_{s_i,i}\ket{\{s_j\}} \nonumber \\
    &=\mathbbm{1} .
\end{align}
Here, the second equality follows from (M4), the third from (M1), the fourth from (\ref{Gsymmprop}), and the fifth from (M2) and (M3). Importantly, the `$\propto$' sign in the second equality is a phase factor that is \emph{independent} of $\{s_j\}$: to see this, note that this phase factor comes from $\bra{\{gs_j\}} (U_g\otimes\mathcal{L}_g) \ket{\{s_j\}} \propto \bra{\{1\}} \mathcal{L}_g \prod_i\G_{g,i}\ket{\{1\}}$ [using (M4)] and both the left and right sides of this identity are clearly independent of $\{s_j\}$.

To complete the argument, we now show that $\W$ is an FDQC (and is therefore a QCA). The first step is to partition the lattice $\Lambda$ into $n$ sublattices $\Lambda_1,\cdots \Lambda_n$ such that the $\G_{g,i}$ operators in each sublattice have non-overlapping regions of support.
 Specifically, if $i,j\in \Lambda_m$ then $\mathrm{supp}(\G_{g,i})\cap \mathrm{supp}(\G_{h,j})=\emptyset$. We then use this partition to rewrite the matrix element $\bra{\{1\}}\prod_i\G_{s_i,i}\ket{\{s_j\}}$ that appears in the definition of $\W$ (\ref{Wdef}):
\begin{align}  \bra{\{1\}} \textstyle \prod_i\G_{s_i,i}\ket{\{s_j\}}&=\bra{\{1\}} \textstyle (\prod_{i\in\Lambda_1}\G_{s_i,i})\cdots (\prod_{i\in\Lambda_n}\G_{s_i,i})\ket{\{s_j\}} \nonumber \\
&=\bra{\{1\}}\textstyle (\prod_{i\in\Lambda_1}\mathcal{R}_{\bar{s}_i,i} \mathcal{V}_{s_i,i})\cdots
(\prod_{i\in\Lambda_n}\mathcal{R}_{\bar{s}_i,i} \mathcal{V}_{s_i,i})\ket{\{s_j\}} \nonumber \\
&=\bra{\{1\}} \textstyle (\prod_{i} \mathcal{R}_{\bar{s}_i,i})(\prod_{i\in\Lambda_1} \tilde{\mathcal{V}}_{\{s_j\},i})\cdots
(\prod_{i\in\Lambda_n} \tilde{\mathcal{V}}_{\{s_j\},i})\ket{\{s_j\}} \nonumber \\
&=\bra{\{s_j\}} \textstyle (\prod_{i\in\Lambda_1} \tilde{\mathcal{V}}_{\{s_j\},i})\cdots
(\prod_{i\in\Lambda_n} \tilde{\mathcal{V}}_{\{s_j\},i})\ket{\{s_j\}} ,
\label{Wfdqc1}
\end{align}
where $\tilde{\mathcal{V}}_{\{s_j\},i}$ is defined by
\begin{equation}
    \tilde{\mathcal{V}}_{\{s_j\},i}=\left(\prod_{j\in I_{m+1}\cup\cdots\cup I_n}\mathcal{R}_{s_j,j}\right)^\dagger\mathcal{V}_{s_i,i}\left(\prod_{j\in I_{m+1}\cup\cdots\cup I_n}\mathcal{R}_{s_j,j}\right)
    \label{Vtildedef},
\end{equation}
for $i\in I_m$. 

A few comments about $\tilde{\mathcal{V}}_{\{s_j\},i}$: first, notice that $\tilde{V}_{\{s_j\},i}$ has the same region of support as $\mathcal{V}_{s_i, i}$ since the two operators are related by conjugation by a product of single site operators. Second, note that $\tilde{\mathcal{V}}_{\{s_j\},i}$ is diagonal in the $\ket{\{s_j\}}$ basis just like $\mathcal{V}_{s_i,i}$. Finally, we need to explain the notation for this operator: the reason that we use the notation $\tilde{\mathcal{V}}_{\{s_j\},i}$ rather than say $\tilde{\mathcal{V}}_{s_i,i}$, is that the operator $\tilde{\mathcal{V}}_{\{s_j\},i}$ doesn't just depend on $s_i$ -- it depends on all the $s_j$'s within its region of support, due to the conjugation in (\ref{Vtildedef}). 

The last step is to define a unitary that combines all the $\tilde{\mathcal{V}}_{\{s_j\},i}$ operators (for fixed $i$) into a single operator, denoted $\hat{\mathcal{V}}_{i}$. Letting $D_i$ denote the region $D_i = \bigcup_{g \in G} \mathrm{supp}(V_{g,i})$, we define
\begin{equation}
\hat{\mathcal{V}}_{i} = \sum_{\{s_j: j \in D_i\}} \tilde{\mathcal{V}}_{\{s_j\},i} P_{D_i}(\{s_j\}),
\end{equation}
where the above sum runs over all possible $\{s_j\}$ configurations within $D_i$, with $P_{D_i}(\{s_j\})$ defined as a projection onto a particular $\{s_j\}$ configuration within $D_i$. By construction, we can replace the $\tilde{\mathcal{V}}_{\{s_j\},i}$ operators in (\ref{Wfdqc1}) by the unitaries $\hat{\mathcal{V}}_{i}$, so that
\begin{align}
\bra{\{1\}} \textstyle \prod_i\G_{s_i,i}\ket{\{s_j\}} = \bra{\{s_j\}} \textstyle (\prod_{i\in\Lambda_1}\hat{\mathcal{V}}_{i})\cdots (\prod_{i\in\Lambda_n}\hat{\mathcal{V}}_{i})\ket{\{s_j\}}    .
\end{align}
Comparing this equation to the definition (\ref{Wdef}), we conclude that
\begin{align}
\W = \left(\prod_{i\in\Lambda_1}\hat{\mathcal{V}}_{i} \right)\cdots \left(\prod_{i\in\Lambda_n}\hat{\mathcal{V}}_{i} \right),
\end{align}
which is the desired FDQC representation of $\W$.

\section{2D $G$-symmetry with general index $[\omega] \in H^2(G, \mathbb{Q}_+)$}

In this section, we construct a general class of 2D anomaly-free, non-onsiteable $G$-symmetries. The construction takes as input a finite group $G$ and an arbitrary 2-cocycle $w:G\times G\to\mathbb{Q}_+$, and outputs a $G$-symmetry with onsiteability index $[\omega]\in H^2(G,\mathbb{Q}_+)$ where $\omega=w^{-1}$. This class of symmetries generalizes the simple $\mathbb{Z}_2$ example introduced in the main text.

Our construction leverages the ``bulk-boundary correspondence" between 3D $G$-QCAs and 2D $G$-symmetries: recall that a ``$G$-QCA" is a QCA that commutes with a set of on-site $G$-symmetry operators $\{O_g\}$. We will first construct a 3D $G$-QCA $V$ parameterized by a cocycle $w:G\times G\to\mathbb{Q}_+$, and then use the bulk-boundary correspondence to define a 2D $G$-symmetry $\{U_g\}$.

\subsection{Map from 3D $G$-QCAs to 2D $G$-symmetries}
\label{sec:bulkboundary}

Here, we review the``bulk-to-boundary" map that produces a 2D $G$-symmetry $\{U_g\}$ given a 3D $G$-QCA $V$. More specifically, we review a particular version of the bulk-to-boundary map that applies in the case that $V$ can be written as an FDQC (whose gates are not necessarily $G$-symmetric).
Consider a large region $A$. Let $N(\partial A)$ denote a neighborhood of the boundary $\partial A$, and $B=\overline{N(\partial A)}$ its complement. Since $V$ can be expressed as an FDQC, there exists a restriction $V_A$ of $V$ to $A$.
The corresponding 2D $G$-symmetry $U_g$, which is supported in $N(\partial A)$, is defined as follows:
\begin{equation}
    U_g=O_{g,B}^\dagger V_A^\dagger O_gV_A
    \label{eq:bulktoboundary}
\end{equation}
where $O_{g,B}=\prod_{i\in B}O_{g,i}$ is the canonical restriction of $O_g$ to $B$. Clearly, $U_g$ is supported within $N(\partial A)$ since $V$ commutes with the global symmetry $O_g$. Moreover, the collection of operators $\{U_g\}$ form a representation of $G$ since $V_A^\dagger O_gV_A$ is a $G$-representation that acts in an on-site fashion outside of $N(\partial A)$, and $O_{g,B}^\dagger$ simply cancels this part of the symmetry action. We note that this bulk-to-boundary map differs from that of \cite{zhang2023}, which applies in the more general setting where $V$ may be a nontrivial QCA even in the absence of $G$-symmetry.

As an aside, we remark that the class of 3D $G$-QCAs constructed in the following subsection are nontrivial $G$-QCAs, meaning they cannot be written as FDQCs composed of individually $G$-symmetric gates. Our construction thus gives a novel class of nontrivial $G$-QCAs beyond SPT entanglers \cite{zhang2023}.

\subsection{A class of nontrivial 3D $G$-symmetric QCA}

We now construct a class of 3D $G$-QCAs parameterized by a 2-cocycle $w:G\times G\to\mathbb{Q}_+$. We note that this construction is reminiscent of the beyond cohomology SPT models described in Ref.~\onlinecite{Gaiotto:2017zba}.

For simplicity, we will present our construction in the case where the cocycle $w:G\times G\to\mathbb{Q}_+$ takes values in the subgroup $\{2^n: n \in \mathbb{Z}\} \subset \mathbb{Q}_+$. This case is particularly simple because we can realize this kind of cocycle using two-state systems, i.e.~\emph{qubits}. At the same time, it is straightforward to generalize from this case to an arbitrary $\mathbb{Q}_+$-valued 2-cocycle: the only new element in the generalization is that one needs to incorporate higher-dimensional \emph{qudit} degrees of freedom in addition to qubits.

As an additional simplification, we will assume that $w$ is a normalized 2-cocycle, i.e.~it obeys the normalization condition
\begin{equation}
    w(1,g)=w(g,1)=1,
    \label{eq:normalization}
\end{equation}
for all $g\in G$. Since any 2-cocycle is gauge-equivalent to a normalized 2-cocycle, we do not lose generality by making this assumption.

We now translate the 2-cocycle $w$ into a slightly different notation that will be convenient below. We do this in two steps. First, we define a 2-cocycle $w':G\times G\to\mathbb{Z}$ by $w'(g,h)=\log_2w(g,h)$.
The 2-cocycle condition on $w'$ reads
\begin{equation}
w'(g,h)+w'(gh,k)=w'(h,k)+w'(g,hk).\label{2cocycleappendix}
\end{equation}
Next we rewrite $w'$ using the ``homogeneous" formulation of group cohomology, in which $w'$ is parameterized by a function $\nu:G\times G\times G\to\mathbb{Z}$ satisfying the homogeneity condition 
\begin{equation}
    \nu(gg_0,gg_1,gg_2)=\nu(g_0,g_1,g_2).
    \label{eq:homogeneity}
\end{equation}
This function is defined via the equality $\nu(g_0,g_1,g_2)=w'(g_0^{-1}g_1,g_1^{-1}g_2)$, and the original cocycle can be recovered via $w'(g,h)=\nu(1,g,gh)$. The 2-cocycle condition (\ref{2cocycleappendix}) on $w'$ implies the following constraint on $\nu$:
\begin{equation}    \nu(g_0,g_1,g_2)+\nu(g_0,g_2,g_3)=\nu(g_1,g_2,g_3)+\nu(g_0,g_1,g_3).\label{eq:homogeneous2cocycle}
\end{equation}

Our construction is defined on an arbitrary 3D spatial manifold $M$ equipped with a triangulation and global vertex ordering. Denote the set of $n$-simplices in the triangulation by $\Delta_n$. We will make use of the dual lattice to the original triangulation. Each site $i$ of the dual lattice corresponds to a 3-simplex of the triangulation, and each edge $ij$ corresponds to a 2-simplex. We use $ij$ to denote both the dual lattice edge and the corresponding 2-simplex.

Consider a Hilbert space of the form $\mathcal{H}=\mathcal{H}_{G\text{-spins}}\otimes\mathcal{H}_\text{qubits}$. Here, $\mathcal{H}_{G\text{-spins}}$ consists of a $G$-valued spin on each vertex $x\in\Delta_0$, with basis states denoted by $\ket{\{g_x\}}$. The on-site $G$-symmetry $\{O_g\}$ acts purely on $\mathcal{H}_{G\text{-spins}}$:
\begin{equation}
    O_g=\sum_{\{g_x\}}\ket{\{gg_x\}}\bra{\{g_x\}}=\prod_{x\in\Delta_0}L^g_x,
    \label{ogdef}
\end{equation}
where $L^g$ is the left multiplication operator.

\begin{figure}[t]
    \centering    \includegraphics[width=.5\linewidth]{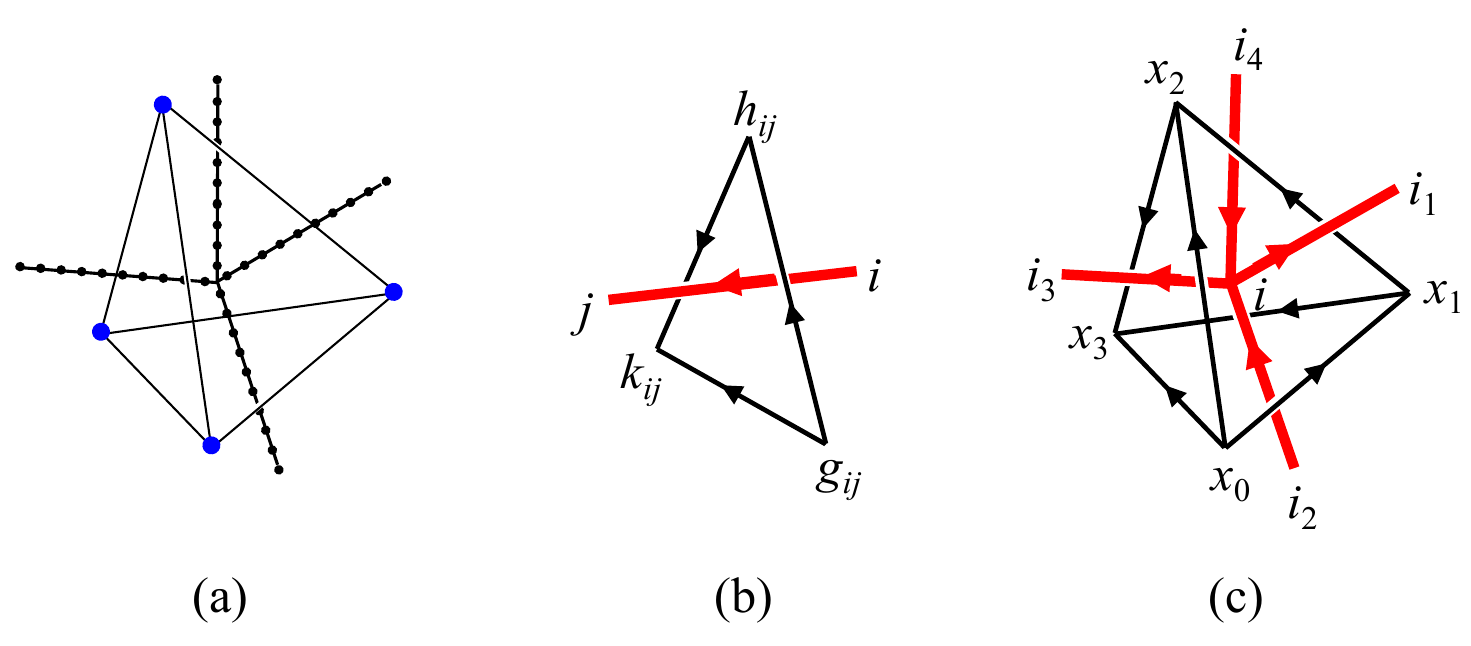}
    \caption{(a) The degrees of freedom of $\mathcal{H}$: a $G$-spin on each vertex of the triangulation (represented by a large blue dot), and a network of 1D qubit chains living on edges of the dual lattice (each small dot represents one qubit). (b) The group elements $g_{ij},h_{ij},k_{ij}$ assigned to the 2-simplex dual to the dual lattice edge $ij$. (c) A 3-simplex in the triangulation of $M$.}
    \label{fig:ghk}
\end{figure}

On the other hand, $\mathcal{H}_\text{qubits}$ consists of a network of 1D qubit chains living on each dual lattice edge $ij\in\Delta_2$, as depicted in Fig.~\ref{fig:ghk}(a). Each of these chains contains a fixed, $\mathcal{O}(1)$ number of qubits that is at minimum larger than $\nu_\text{max}=\max_{g_i}|\nu(g_0,g_1,g_2)|$. There are thus two relevant length scales in this setup: the ``microscopic" 1D qubit chains, and the ``mesoscopic" lattice of the 3-manifold triangulation. We regard a unitary operator to be a QCA if it is locality preserving with respect to the mesoscopic length scale.

Our 3D $G$-QCA, denoted by $V$, acts as a global permutation of the qubits in $\mathcal{H}_\text{qubits}$, controlled by the values of the $G$-spins in $\mathcal{H}_{G\text{-spins}}$. More specifically, given a collection of group elements $\{g_x\}$, each dual lattice edge $ij$ is assigned an integer $\nu_{ij}=\nu(g_{ij},h_{ij},k_{ij})$, where $g_{ij},h_{ij},k_{ij}$ are the group elements assigned to the vertices of the 2-simplex dual to $ij$ (see Fig.~\ref{fig:ghk}(b)). By convention $\nu_{ji}=-\nu_{ij}$. Due to the 2-cocycle condition (\ref{eq:homogeneous2cocycle}) on $\nu$, the set of integers $\{\nu_{ij}\}$ is guaranteed to satisfy the ``divergence-free" constraint $\nu_{i,i_1}+\nu_{i,i_3}=\nu_{i_2,i}+\nu_{i_4,i}$, where $i_1,i_2,i_3,i_4$ are the 3-simplices adjacent to $i$ (see Fig.~\ref{fig:ghk}(c)). The $G$-QCA $V$ is defined as follows:
\begin{equation}
    V=\hat{T}[\{\nu_{ij}\}],
\end{equation}
where $\hat{T}$ is a functional whose input is the set of integers $\{\nu_{ij}\}$, and whose output is a unitary operator on $\mathcal{H}_\text{qubits}$ that simply permutes the set of qubits. The basic idea is that $\nu_{ij}$ quantifies the flow of qubits between $i$ and $j$. If $\nu_{ij}>0$, then there are $\nu_{ij}$ qubits flowing from $i$ to $j$, and if $\nu_{ij}<0$, then there are $|\nu_{ij}|$ qubits flowing from $j$ to $i$. The divergence-free condition on $\{\nu_{ij}\}$ thus states that there is a net zero flow of qubits into/out of each 3-simplex.

To make this precise, we need to specify where the permutation sends each individual qubit. Within the bulk of each dual lattice edge, the action of $\hat{T}[\{\nu_{ij}\}]$ is straightforward: it acts as a translation of the 1D qubit chain by $\nu_{ij}$ sites. It remains to specify how $\hat{T}[\{\nu_{ij}\}]$ acts in the vicinity of the dual lattice vertices, each of which lies at the junction of four 1D qubit chains. There are $N_i=\nu_{i,i_1}+\nu_{i,i_3}=\nu_{i_2,i}+\nu_{i_4,i}$ qubits flowing through the dual lattice vertex $i$. The outgoing $N_i$ qubits are mapped by translation further along their respective chains, and the incoming $N_i$ qubits take their place. We choose an arbitrary convention for the order in which incoming qubits fill the outgoing vacancies. The specific choice of convention is not important: the equivalence class of the $G$-QCA $V$, and the onsiteability index of the 2D $G$-symmetry derived below, are fully independent of this choice.

Finally, we note that $V$ is manifestly $G$-symmetric since the configuration $\{\nu_{ij}\}$ is itself invariant under the action of $O_g$, due to the homogeneity condition (\ref{eq:homogeneity}).

\subsection{Derivation of the 2D $G$-symmetry}
\label{sec:2D}

\begin{figure}[t]
    \centering    \includegraphics[width=0.25\linewidth]{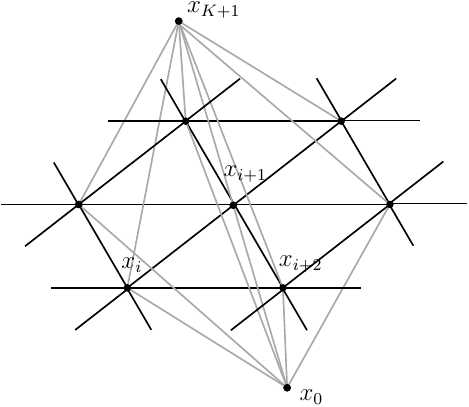}
    \caption{A triangulated surface $X$ embedded within the triangulated $3$-manifold $SX$. Each vertex in $X$ is labeled $x_i$, where $i\in\{1,\cdots,K\}$. The vertices at the north and south poles are labeled $x_0$ and $x_{K+1}$, respectively.}
    \label{fig:3_manifold}
\end{figure}

In this subsection, we will use the bulk-to-boundary map from Sec.~\ref{sec:bulkboundary} to construct a 2D $G$-symmetry based on the $G$-QCA $V$ defined above. We begin by describing the geometric setup of the construction.
Our $G$-symmetry will act on a 2D surface $X$ without boundary, which we equip with a triangulation and a global ordering on the set of vertices, $\Delta_0^X=\{x_1,\ldots,x_K\}$. We denote the 1-simplices and 2-simplices of the triangulation by $\Delta_1^X$ and $\Delta_2^X$. In order to use the bulk-to-boundary map, we will embed the surface $X$ within the 3-manifold $SX$, the suspension of $X$. Recall that $SX$ is defined as the quotient space $(X\times[0,1])/(X\times\{0\})/(X\times\{1\})$, which can be thought of as a ``northern hemisphere" and a ``southern hemisphere" glued together along $X$, the ``equator". We extend the 2D triangulation of $X$ to a 3D triangulation of $SX$ by adding a ``north pole" vertex $x_{K+1}$ and a ``south pole" vertex $x_0$, together with edges connecting each pole to each vertex of the original triangulation. The vertices of $SX$ are globally ordered, $\Delta_0=\{x_0,x_1,\ldots,x_{K+1}\}$. The set of 3-simplices $\Delta_3$ is naturally divided into two subsets, denoted by $\Delta_3^N$ and $\Delta_3^S$, which comprise the northern and southern hemispheres respectively. Each of these subsets is in one-to-one correspondence with $\Delta_2^X$. On the other hand, the set of 2-simplices $\Delta_2$ consists of three subsets, $\Delta_2^N,\Delta_2^S$, and $\Delta_2^X$. The subsets $\Delta_2^N$ and $\Delta_2^S$ are in one-to-one correspondence with $\Delta_1^X$. For reference, see Fig.~\ref{fig:3_manifold}.

Next, we apply the construction of the previous subsection to the triangulated manifold $SX$. Recall that this construction produces a $G$-QCA $V$ acting on the Hilbert space $\mathcal{H}=\mathcal{H}_{G\text{-spins}}\otimes\mathcal{H}_\text{qubits}$, where $\mathcal{H}_{G\text{-spins}}$ consists of a $G$-valued spin on each vertex $x\in\Delta_0$, and $\mathcal{H}_\text{qubits}$ consists of a 1D chain of qubits on each edge of the dual lattice. This system has the on-site $G$-symmetry $O_g$ defined in Eq.~(\ref{ogdef}).

We are now ready to apply the ``bulk-to-boundary" mapping. Let $A$ be the northern hemisphere, and $B$ be a neighborhood of the union of the two poles, such that the restricted symmetry $O_{g,B}=L^g_{x_0}L^g_{x_{K+1}}$. Furthermore, we choose the following restriction $V_A$ of $V$ to $A$:
\begin{equation}
    V_A=\hat{T}[\{\tilde\nu_{ij}\}],
\end{equation}
where
\begin{equation}
    \tilde\nu_{ij}=\begin{cases}
		\nu(1,h_{ij},k_{ij}) & \text{if $ij\in \Delta_2^S$}\\
            \nu(g_{ij},h_{ij},k_{ij}) & \text{otherwise}.
		 \end{cases}
\end{equation}
Note that if $ij\in\Delta_2^S$, then $g_{ij}$ is the group element assigned to the south pole vertex $x_0$, and is otherwise assigned to a vertex $x\in\Delta_0^X$. Thus, $V_A$ is indeed a restriction of $V$ to $A$, since its support excludes the $G$-spin on the south pole. However, it does act on the qubits living in the southern hemisphere.

The 2D $G$-symmetry operators are defined according to (\ref{eq:bulktoboundary}):
\begin{equation}
    U_g=O_{g,B}^\dagger V_A^\dagger O_gV_A.
\end{equation}
A simple computation shows that
\begin{equation}
    U_g=O_{g,X}\hat{T}[\{\hphantom{}^g\tilde\nu_{ij}\}]^\dagger\hat{T}[\{\tilde\nu_{ij}\}]
    \label{eq:Ug},
\end{equation}
where $O_{g,X}=\prod_{x\in\Delta_0^X}L^g_x$ is the restriction of $O_g$ to $X$, and
\begin{equation}
    ^g\tilde\nu_{ij}=\begin{cases}
		\nu(1,gh_{ij},gk_{ij}) & \text{if $ij\in \Delta_2^S$}\\
            \nu(gg_{ij},gh_{ij},gk_{ij}) & \text{otherwise.}
            
		 \end{cases}.
\end{equation}

\begin{figure}[t]
    \centering    \includegraphics[width=\linewidth]{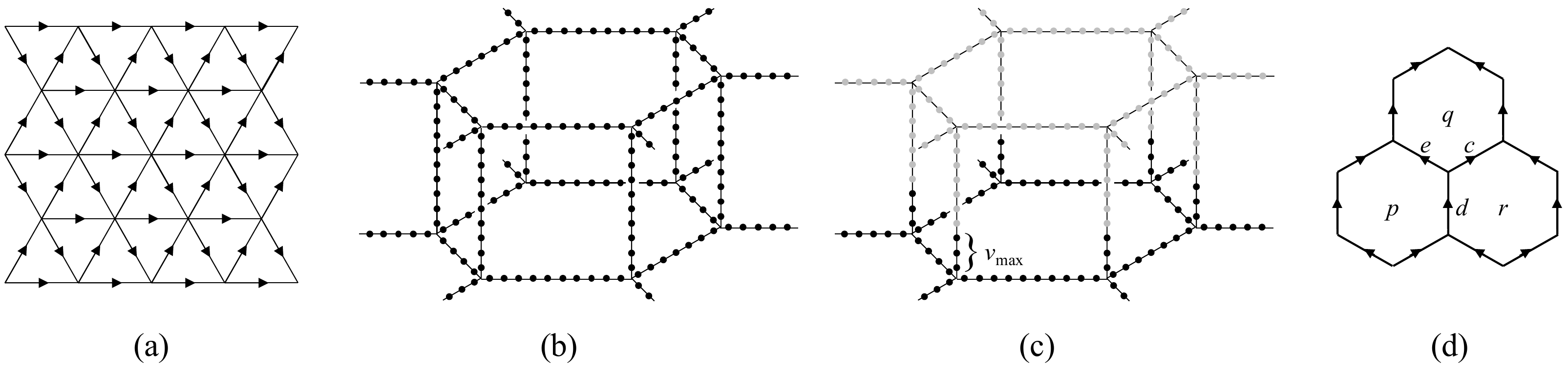}
    \caption{(a) Triangular lattice with a global vertex ordering. (b) 1D qubit chains living on edges of the dual lattice. Each dot represents a qubit. (c) The subset of qubits that comprise the support of the 2D symmetry operator $U_g$. The gray dots represent qubits acted on trivially by $U_g$. (d) Plaquettes $p,q,r$ and edges $c,d,e$ of the honeycomb lattice.}
    \label{fig:torus}
\end{figure}

While (\ref{eq:Ug}) is a valid expression, it is not immediately clear that $U_g$ acts on a strictly 2D subsystem. To see this, note that $\hat{T}[\{\hphantom{}^g\tilde\nu_{ij}\}]^\dagger\hat{T}[\{\tilde\nu_{ij}\}]$ acts trivially on all qubits lying on dual lattice edges $ij\in\Delta_2^X,\Delta_2^N$ (with the possible exception of the $\nu_\text{max}$ qubits at the southern end edges in $\Delta_2^X$ -- see Fig.~\ref{fig:torus}(c)), since $^g\tilde\nu_{ij}=\tilde\nu_{ij}$ for all $ij\in\Delta_2^X,\Delta_2^N$ due to the homogeneity condition (\ref{eq:homogeneity}). Therefore, the action of $\hat{T}[\{\hphantom{}^g\tilde\nu_{ij}\}]^\dagger\hat{T}[\{\tilde\nu_{ij}\}]$ is independent of the $G$-spin on the north pole. As this action is also manifestly independent of the $G$-spin on the south pole, we conclude that $U_g$ acts trivially on both poles.

In other words, we may express the symmetry operators as
\begin{equation}
    U_g=O_{g,X}\hat{T}[\{\hphantom{}^g\tilde{\tilde\nu}_{ij}\}]^\dagger\hat{T}[\{\tilde{\tilde\nu}_{ij}\}]
    \label{eq:Ug},
\end{equation}
where
\begin{equation}
    \tilde{\tilde\nu}_{ij}=\begin{cases}
		\nu(1,h_{ij},k_{ij}) & \text{if $ij\in \Delta_2^S$}\\
            \nu(g_{ij},h_{ij},k_{ij}) & \text{if $ij\in \Delta_2^X$}\\
            \nu(g_{ij},h_{ij},1) & \text{if $ij\in \Delta_2^N$}
            
		 \end{cases},\qquad
         ^g\tilde{\tilde\nu}_{ij}=\begin{cases}
		\nu(1,gh_{ij},gk_{ij}) & \text{if $ij\in \Delta_2^S$}\\
            \nu(gg_{ij},gh_{ij},gk_{ij}) & \text{if $ij\in \Delta_2^X$}\\
            \nu(gg_{ij},gh_{ij},1) & \text{if $ij\in \Delta_2^N$}
            
		 \end{cases}.
\end{equation}
As a check, let us verify the group multiplication law:
\begin{align}
    U_gU_h&=O_{g,X}\hat{T}[\{\hphantom{}^g\tilde{\tilde\nu}_{ij}\}]^\dagger\hat{T}[\{\tilde{\tilde\nu}_{ij}\}]O_{h,X}\hat{T}[\{\hphantom{}^h\tilde{\tilde\nu}_{ij}\}]^\dagger\hat{T}[\{\tilde{\tilde\nu}_{ij}\}] \nonumber \\
    &=O_{g,X}O_{h,X}\hat{T}[\{\hphantom{}^{gh}\tilde{\tilde\nu}_{ij}\}]^\dagger\hat{T}[\{\hphantom{}^h\tilde{\tilde\nu}_{ij}\}]\hat{T}[\{\hphantom{}^h\tilde{\tilde\nu}_{ij}\}]^\dagger\hat{T}[\{\nu_{ij}\}] \nonumber \\
    &=O_{gh,X}\hat{T}[\{\hphantom{}^{gh}\tilde{\tilde\nu}_{ij}\}]^\dagger\hat{T}[\{\tilde{\tilde\nu}_{ij}\}] \nonumber \\
    &=U_{gh}.
\end{align}

To illustrate the construction, consider the case where the 2D surface $X$ is a torus triangulated by a standard triangular lattice, as shown in Fig.~\ref{fig:torus}(a). In this case, the dual vertices of $SX$ form two honeycomb lattices, one in the northern hemisphere, and one in the southern hemisphere. The 1D qubit chains comprising $\mathcal{H}_\text{qubits}$ live on the edges of the dual lattice, as depicted in Fig.~\ref{fig:torus}(b). The symmetry operators $U_g$ are supported on the subsystem $\mathcal{H}'=\mathcal{H}_{G\text{-spins}}'\otimes\mathcal{H}_\text{qubits}'$, where $\mathcal{H}'_{G\text{-spins}}\subset\mathcal{H}_{G\text{-spins}}$ consists of all $G$-spins in $X$, and $\mathcal{H}'_\text{qubits}\subset\mathcal{H}_\text{qubits}$ is composed of a subset of the qubits living in the southern hemisphere (see Fig.~\ref{fig:torus}(c)).

Thus, we can view $\mathcal{H}'$ as living on a purely 2D honeycomb lattice, with a single $G$-spin per plaquette $p$, a 1D chain of qubits on each edge, and an additional set of qubits (of quantity $\nu_\text{max}$) on each vertex. The symmetry operators may be expressed in the form
\begin{equation}
    U_g=O_g\hat{S}_g[\{g_p\}],
    \label{gensymmdef}
\end{equation}
where $O_g=\prod_pL^g_p$, and $\hat{S}_g[\{g_p\}]$ is a permutation of the qubits in $\mathcal{H}'_\text{qubits}$ controlled by the $G$-spins in $\mathcal{H}'_{G\text{-spins}}$. The key property of this functional is the following: within the bulk of each edge $e$, the unitary $\hat{S}_g[\{g_p\}]$ acts as a pure translation of the 1D qubit chain by $m$ sites where
\begin{equation}
    m=\nu(1,g_p,g_q)-\nu(1,gg_p,gg_q).\label{eq:m}
\end{equation}
Here $p$ and $q$ are the plaquettes adjacent to $e$.

To be precise, consider the plaquettes $p,q,r$ and edges $c,d,e$ shown in Fig.~\ref{fig:torus}(d). The unitary $\hat{S}_g[\{g_p\}]$ acts as a translation by $m_e=\nu(1,g_p,g_q)-\nu(1,gg_p,gg_q)$ sites along $e$, by $m_c=\nu(1,g_q,g_r)-\nu(1,gg_q,gg_r)$ sites along $c$, and by $m_d=\nu(1,g_p,g_r)-\nu(1,gg_p,gg_r)$ sites along $d$ (in the direction of the arrows). As a check, let us verify that the number of incoming qubits is the same as the number of outgoing qubits. Indeed, from the homogeneity condition (\ref{eq:homogeneity}) and the 2-cocycle condition (\ref{eq:homogeneous2cocycle}), it follows that
\begin{equation}
    m_e+m_c-m_d=\nu(1,g_p,g_q)-\nu(1,gg_p,gg_q)+\nu(1,g_q,g_r)-\nu(1,gg_q,gg_r)-\nu(1,g_p,g_r)+\nu(1,gg_p,gg_r)=0.
\end{equation}

\subsection{Example: $G=\mathbb{Z}_2$}

Here we illustrate the construction via a simple example with $G=\mathbb{Z}_2=\{1,g\}$. The cohomology group $H^2(\mathbb{Z}_2,\mathbb{Z})=\mathbb{Z}_2$, where the nontrivial class has representative cocycle $w':\mathbb{Z}_2\times\mathbb{Z}_2\to\mathbb{Z}$ with $w'(g,g)=1$ and $w'(1,g)=w'(g,1)=w'(1,1)=0$. Hence $\nu(1,g,1)=\nu(g,1,g)=1$, and $\nu=0$ for all other sets of arguments. In the 2D honeycomb lattice system, the $G$-spin on each plaquette is simply a qubit, where the $\ket{1}$ state is identified with $\ket{\up}$ and the $\ket{g}$ state with $\ket{\down}$. Within the bulk of each edge, the 2D symmetry $U_g$ acts as a translation by $\nu(1,h,k)-\nu(1,gh,gk)$ on the 1D qubit chain, where $h$ and $k$ are the group elements on the adjacent plaquettes. This amounts to a translation by one site along each domain wall, where the $\ket{\down}$ spin is to the left when facing the direction of forward translation.

This symmetry operator is closely related to the $\mathbb{Z}_2$ symmetry discussed in the main text. The example in the main text is a `minimalist' version in which the 1D chains of qubits on edges of the honeycomb are replaced by single qubits on each vertex.

\subsection{Onsiteability index calculation}

We now compute the onsiteability index $[\omega]\in H^2(G,\mathbb{Q}_+)$ of the 2D $G$-symmetry $\{U_g\}$ derived in the previous subsection. As before, we specialize to the case where $X$ is a torus triangulated by a standard triangular lattice. For this class of examples, it is sufficient to consider the disk $A=q$, where $q$ is a single plaquette of the honeycomb lattice. We choose the following restriction $U_{g,A}$ of each symmetry operator $U_g$ to the vicinity of plaquette $q$:
\begin{equation}
    U_{g,A}=O_{g,A}\hat{S}_g[\{\tilde g_p\}],
\end{equation}
where
\begin{equation}
    \tilde g_p=\begin{cases}
        g_q &\text{if $p=q$}\\
        1&\text{otherwise.}
    \end{cases}
\end{equation}
To compute the onsiteability index $[\omega]$, we need only compute the GNVW index of the operator \begin{align}
    \Omega_{g,h}&=U_{g,A}U_{h,A}U_{gh,A}^{-1} \nonumber \\
    &=O_{g,A}\hat{S}_g[\{\tilde g_p\}]O_{h,A}\hat{S}_h[\{\tilde g_p\}]\hat{S}_{gh}[\{\tilde g_p\}]^\dagger O_{gh,A}^\dagger.\label{eq:Omegagh}
\end{align}
To do so, we exploit the fact that the GNVW index is defined locally. Thus, it is sufficient to consider the action of $\Omega_{g,h}$ within the bulk of a single edge $e\in q$. As we will show, this action is particularly simple: it is a pure translation of the 1D qubit chain on $e$.

To see this, consider the geometry in Fig.~\ref{fig:torus}(d). Within the bulk of $e$, the symmetry $U_g$ acts as a translation by a number of sites controlled by the $G$-spins on $p,q$. Thus, each $U_{g,A}$ acts as a translation by a number of sites controlled purely by the $G$-spin on $q$. \emph{A priori}, $\Omega_{g,h}$ also acts as a translation by a number of sites controlled by the $G$-spin on $q$. However, it turns out that $\Omega_{g,h}$ is in fact independent of this $G$-spin. To see this, take an arbitrary group element $k\in G$, and consider the operator $\bra{k}_q\Omega_{g,h}\ket{k}_q$ acting on $\mathcal{H}_\text{qubits}'$. Within the bulk of $e$, this operator is a translation by $n$ sites, where
\begin{align}
    n&=\nu(1,1,hk)-\nu(1,g,k)+\nu(1,1,(gh)^{-1}k)-\nu(1,h,g^{-1}k)-\nu(1,1,(gh)^{-1}k)+\nu(1,gh,k) \nonumber \\
    &=-\nu(1,g,k)-\nu(g,gh,k)+\nu(1,gh,k) \nonumber \\
    &=-\nu(1,g,gh) \nonumber \\
    &=-w'(g,h).
\end{align}
Here, the first equality is obtained by substituting (\ref{eq:m}) into the expression (\ref{eq:Omegagh}). The second equality follows from the normalization condition (\ref{eq:normalization}) and the homogeneity condition (\ref{eq:homogeneity}), and the third equality follows from the 2-cocycle constraint (\ref{eq:homogeneous2cocycle}). Since $n$ is independent of $k$, we conclude that $\Omega_{g,h}$ acts as a pure translation by $n$ sites on the qubits within the bulk of $e$. Thus, we find that $\mathrm{Ind}\left(\Omega_{g,h}\right)=w(g,h)^{-1}$, hence $[\omega]=[w^{-1}]$.

\subsection{Trivial ground state and parent Hamiltonian}
We now show that the symmetries defined above are anomaly-free. We do this by constructing $U_g$-symmetric product states with gapped parent Hamiltonians. To begin, notice that the $U_g$ symmetries (\ref{gensymmdef}) have a very similar form to the $\mathbb{Z}_2$ example in the main text: the qubits are permuted according to the plaquette ($G$-spin) domain walls. Therefore the symmetry $U_{g}$ admits the symmetric product state
\begin{equation}
    |\Psi_0\rangle=\otimes_{p,i}|L_p^g=+1,\tau^x_i=+1\rangle,
\end{equation}
where the qubits $\tau^x_i$ form the 1D chains along the edges. To see why $|\Psi_0\rangle$ is invariant under $U_g$, notice that $\hat{S}_g[\{g_p\}]$ performs permutations on the qubits, and therefore does nothing to $|\Psi_0\rangle$ since the qubits are in a translation invariant state. Clearly $|\Psi_0\rangle$ is also invariant under $O_g$, so it is invariant under $U_g=O_g\hat{S}_g[\{g_p\}]$.

Like for the $\mathbb{Z}_2$ example in the main text, we can write down a frustration-free, symmetric parent Hamiltonian whose unique ground state is $|\Psi_0\rangle$:
\begin{equation}\label{hamiltonian}
    H=-\sum_i\tau^x_i-\sum_{p,g,h}U_h^{-1}L^g_pU_h,
\end{equation}
where we sum over all group elements $h\in G$ in order to symmetrize $L_p^g$. Note that the first term $\sum_i\tau^x_i$ is by itself already symmetric, because it is translation invariant. The Hamiltonian $H$ is not commuting, but it is gapped for the same reasons as discussed in the main text for the $\mathbb{Z}_2$ example. Specifically, since $[H,\sum_i\tau^x_i]=0$, we can diagonalize $H$ separately within each of the different eigenspaces of $\sum_i\tau_i^x$. We actually only need to consider the sector $\sum_i\tau^x_i=N_{\text{site}}$, the maximally polarized sector, which contains $|\Psi_0\rangle$. This is because all other sectors are higher in energy by at least $2$ due to the first term in $H$. Within this subspace, $H$ reduces to $H = - N_{\text{site}} - |G|\sum_{p,g}L_p^g$, so the unique gapped ground state is $|\Psi_0\rangle$.

\section{Example: anyon-permuting symmetry of the toric code}

In this section, we discuss an anomaly-free, non-onsiteable $\mathbb{Z}_4$ symmetry of the square lattice toric code that permutes the $e$ and $m$ anyons, which was discovered originally in Ref.~\onlinecite{Barkeshli:2022wuz} (see \cite{radical} for a similar model). Beyond the fact that it permutes anyons, this example is notable because the symmetry is a circuit composed of Clifford gates. It therefore has a particularly simple action on operators: it maps every Pauli string to another Pauli string.

\subsection{Definition of the $\mathbb{Z}_4$ symmetry}
\begin{figure}[tb]
   \centering   \includegraphics[width=.25\columnwidth]{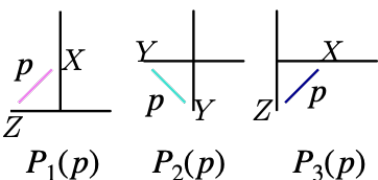} 
   \caption{$U_g$ is a depth-three circuit built out of the Pauli strings $P_1(p)$, $P_2(p)$, and $P_3(p)$ associated with every plaquette $p$, illustrated above.}
   \label{fig:bosonization}
\end{figure}

\begin{figure}[tb]
   \centering   \includegraphics[width=.5\columnwidth]{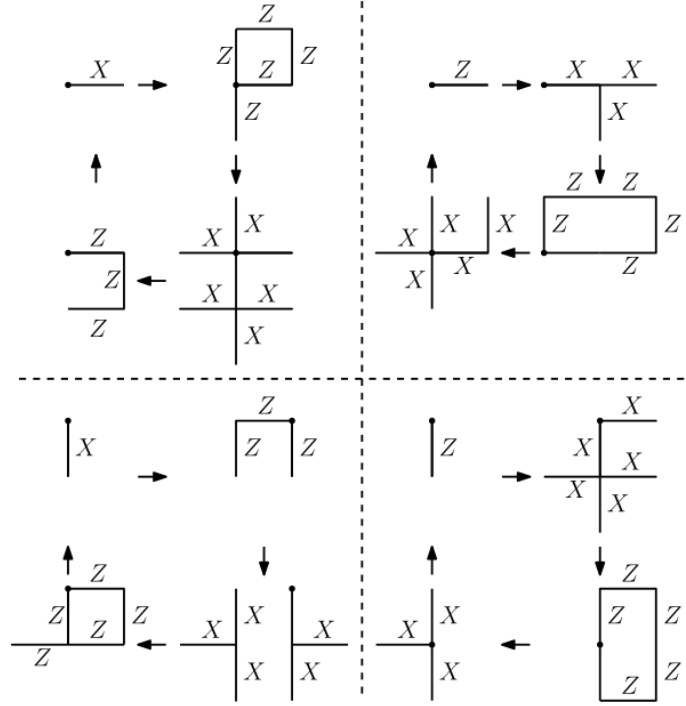} 
   \caption{$U_g$ has a $\mathbb{Z}_4$ action on all operators. Its action on any operator can be deduced from its action on the Pauli operators. In each action, a Pauli string is mapped as shown by an arrow, together with a minus sign. The black dots anchor the locations of the initial Pauli operators.}
   \label{fig:z4action}
\end{figure}

The toric code Hilbert space is composed of a single qubit on each edge of a square lattice. The Hamiltonian takes the form
\begin{equation}\label{tcham}
    H=-\sum_vA_v-\sum_pB_p,
\end{equation}
where $v$ ($p$) indexes the vertices (plaquettes) of the lattice. Here $A_v$ is a product of Pauli $X$ operators over the four edges adjacent to $v$, and $B_p$ is a product of Pauli $Z$ operators over the four edges comprising $p$.
The generator of the $\mathbb{Z}_4=\{1,g,g^2,g^3\}$ symmetry can be described as a depth-three circuit where each layer is composed of two-qubit Clifford gates. In particular, it takes the form
\begin{equation}\label{z4symm}
    U_g=\prod_pe^{i\pi P_3(p)/4}\prod_pe^{i\pi P_2(p)/4}\prod_pe^{i\pi P_1(p)/4},
\end{equation}
where $P_1(p),P_2(p),P_3(p)$ are the two-qubit Pauli strings shown in Fig.~\ref{fig:bosonization}. Recall that for any Pauli string $P$, the unitary $U=e^{i\pi P/4}$ is a Clifford gate with the following action on a generic Pauli string $Q$:
\begin{equation}
    UQU^\dagger=\begin{cases}
        Q&\text{if $[P,Q]=0$}\\
        iPQ&\text{if $\{P,Q\}=0$}.
    \end{cases}
    \label{eq:Clifford_gate}
\end{equation}
This fact can be used to deduce the action of $U_g$ on local operators, as shown in Fig.~\ref{fig:z4action}. It is straightforward to check that
\begin{equation}
    U_gA_vU_g^\dagger=B_p,\quad U_gB_pU_g^\dagger=A_v,
\end{equation}
where $v$ is the bottom left vertex of plaquette $p$. Thus $U_g$ commutes with the toric code Hamiltonian $H$, and permutes the $e$ and $m$ anyons which are excitations of the vertex and plaquette terms, respectively. Note that unlike the familiar $e-m$ permuting symmetry of the square lattice toric code (which consists of a Hadamard gate swapping $Z\leftrightarrow X$ on every link combined with a diagonal translation) this symmetry does not mix with translation.

\subsection{Calculation of the index}\label{calcindex}
We will now compute the onsiteability index $[\omega]\in H^2(\mathbb{Z}_4,\mathbb{Q}_+)$ of the symmetry generated by $U_g$.
The class $[\omega]\in H^2(\mathbb{Z}_4,\mathbb{Q}_+)$
is fully characterized by the quantity
\begin{equation}
    \Theta=\omega(1,g)\omega(g,g)\omega(g^2,g)\omega(g^3,g).\label{eq:Theta}
\end{equation}
This quantity has a simple physical interpretation: given a restriction $U_{g,A}$ of $U_g$ to a disk $A$, the number $\Theta$ corresponds to the GNVW index of $U_{g,A}^4$, which is a 1D QCA supported near $\partial A$. To see this, note that
\begin{align}
    \Theta=\mathrm{Ind}\Big(U_{1,A}U_{g,A}U_{g,A}^{-1}\Big)\cdot\mathrm{Ind}\Big(U_{g,A}U_{g,A}U_{g^2,A}^{-1}\Big)\cdot\mathrm{Ind}\Big(U_{g^2,A}U_{g,A}U_{g^3,A}^{-1}\Big)\cdot\mathrm{Ind}\Big(U_{g^3,A}U_{g,A}U_{1,A}^{-1}\Big)
    =\mathrm{Ind}\Big(U_{g,A}^4\Big),\label{eq:Theta_derivation}
\end{align}
where we have used the fact that the GNVW index is multiplicative under composition of operators. Here $U_{1,A},U_{g^2,A},U_{g^3,A}$ are arbitrary restrictions of $U_1,U_{g^2},U_{g^3}$ respectively. Note that a shift of $\omega(g,h)\to \omega(g,h)\cdot \alpha(g)\alpha(h)/\alpha(gh)$ multiplies $\Theta$ by $\alpha(g)^4$, which is a fourth power of a rational number. 

\begin{figure}[tb]
   \centering   \includegraphics[width=.6\columnwidth]{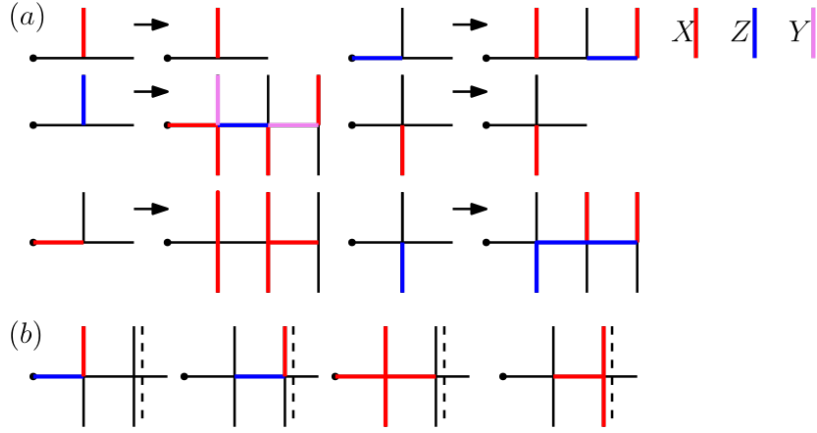} 
   \caption{$(a)$ The action of $U_{g, A}^4$ on single-site Pauli operators supported near the top boundary of $A$, where we used red, blue, and purple to indicate Pauli $X,Z,$ and $Y$ operators respectively (see the legend to the right of the figure). The black dots anchor the locations of the initial Pauli operators. One can check that conjugation by $U_{g,A}^4$ maintains all commutation relations, as expected. $(b)$ These four operators generate 16 operators that are completely transported across the dashed line by $U_{g,A}^4$, giving a GNVW index of $\sqrt{16}=4$.} 
   \label{fig:bosonicedge}
\end{figure}

Recall that $U_{g,A}^4$ leaves all Pauli operators in the bulk of $A$ invariant. We show in Fig.~\ref{fig:bosonicedge}(a) the action of $U_g^4$ on all Pauli operators supported near the boundary of $A$, with a particular choice of truncation. In Fig.~\ref{fig:bosonicedge}(b) we show that for a given cut along the boundary of $A$, 16 operators get transported by $U_{g,A}^4$ from having full support to the left of the cut to having full support to the right of the cut. No operators get transported in the opposite direction. This result is independent of the choice of cut, and results in a GNVW index of 4 according to the equation for the GNVW index in Sec.~\ref{GNVWreview}. It follows that $\Theta=4$, which is not the fourth power of any rational number, so $[\omega]$ is a nontrivial element of $H^2(\mathbb{Z}_4,\mathbb{Q}_+)$.

\section{Fermionic examples}
In this section, we discuss two examples of fermionic symmetries that are anomaly-free and not onsiteable. Our first example is notable because the symmetry is particularly simple: it is a Gaussian operator of the form $U = e^{i\sum_{jk} A_{jk} c_j^\dagger c_k}$. Our second example has the interesting property that the symmetry transformation provides a (non-symmetric) entangler for the two dimensional fermionic root SPT phase with symmetry group $\mathbb{Z}_2 \times \mathbb{Z}_2^f$. 


\subsection{Non-onsiteable Gaussian symmetries}

One of the simplest kinds of unitary symmetry transformations that can act in a many-body Fock space are \emph{Gaussian} operators of the form $U = e^{i\sum_{jk} A_{jk} c_j^\dagger c_k}$. In this section, we construct anomaly-free, non-onsiteable fermionic symmetries of this kind.

We begin by explaining our setup. We consider spinless fermions hopping on a 2D lattice $\Lambda$. We denote the fermion creation and annihilation operators on site $j \in \Lambda$ by $c_j^\dagger, c_j$. These operators satisfy the usual anticommutation relations: $\{c_j, c_k^\dagger\} = \delta_{jk}$ and $\{c_j, c_k\} = 0$. The basic input needed to construct one of our symmetries is a Hermitian projection matrix $P_{jk}$ that is ``quasi-diagonal'' in the indices $j, k \in \Lambda$ in the sense that its matrix elements decay superpolynomially away from the diagonal: $|P_{jk}|\leq \mathcal{O}\left(|j-k|^{-\infty}\right)$ \footnote{Ref.~\cite{kitaev2006} also considered slower decay, but in this work we will assume that the off-diagonal matrix elements decays superpolynomially.} Any such $P$ has a  well-defined Chern number $\nu(P) \in \mathbb{Z}$ \cite{kitaev2006}; for our construction, we require that $\nu(P) \neq 0$.

Given a Chern projector $P$ with the above properties, we define a corresponding $\mathbb{Z}_{n}$ symmetry transformation by
\begin{align}
U_{g} = e^{2\pi i/n \sum_{jk} P_{jk} c_j^\dagger c_k}.
\label{Gausssym}
\end{align}
Notice that $U_g^{n} = \mathbbm{1}$, as required for a $\mathbb{Z}_{n}$ symmetry, since $P$ has eigenvalues $0$ and $1$. Also, notice that the unitary $U_g$ is locality preserving: $U_g c_j U_g^\dagger$ is supported in a neighborhood of $j$, with superpolynomially decaying tails (since $P$ is quasi-diagonal). At the same time, $U_g$ does not satisfy the \emph{strict} locality preservation properties of a QCA, due to the fact that Chern projectors can never be strictly finite range\cite{chen2014}.

A related issue with the $\mathbb{Z}_{n}$ symmetry (\ref{Gausssym}) is that $U_g$ cannot be written exactly as an FDQC. Therefore, strictly speaking the definition of the onsiteability index given in the main text does not apply to $U_g$. That said, we expect that our onsiteability index can be generalized to the larger class of ``locally generated'' symmetries which includes $U_g$. Here, a locally generated symmetry is one that can be written in the form $U = \mathcal{T} \exp( -i \int_0^T dt H(t))$ where $H(t)$ is a local Hamiltonian, possibly with superpolynomially decaying tails. Such locally generated unitaries are close analogs of FDQCs and share many of the same basic properties, so we will assume that the formalism from the main text can be extended to these symmetries. It is also clear that $U_g$ (\ref{Gausssym}) is locally generated since we can choose $H = -\sum_{jk} P_{jk} c_j^\dagger c_k$. 

We now show that the $\mathbb{Z}_{n}$ symmetry (\ref{Gausssym}) is anomaly-free. We do this by constructing an explicit gapped, symmetric Hamiltonian with a product state as its unique ground state. In particular, consider the Hamiltonian $H = \sum_j c_j^\dagger c_j$, which is simply the total particle number. Clearly $[H, U_g] = 0$. Also, $H$ is gapped and has the vacuum state $\ket{\{c_j^\dagger c_j = 0\}}$ as its unique ground state. Hence the symetry (\ref{Gausssym}) is anomaly-free according to the definition given in the main text.

At the same time, the $\mathbb{Z}_{n}$ symmetry (\ref{Gausssym}) is non-onsiteable if the Chern number $2\nu(P)$ is not divisible by $n$. To see this, let us restrict $U_g$ to a half-plane $A$, defining
\begin{align}
    U_{g,A} = e^{2\pi i/n \sum_{jk} P_{A,jk} c_j^\dagger c_k},
\end{align}
where 
\begin{align}
    P_{jk,A}=\begin{cases} P_{jk} & j,k\in A\\ 0 & \text{otherwise}. \end{cases}
\end{align}
Next, consider the operator $U_{g,A}^{n}$, which is supported near $\partial A$. We claim that the GNVW index of this operator is given by
\begin{align}
   \mathrm{Ind}(U_{g,A}^{n}) = 2^{\nu(P)} .
   \label{indexcal}
\end{align}
This result, which we derive below, has immediate implications for the non-onsiteability of $U_g$. To see this, recall that $\mathrm{Ind}(U_{g,A}^{n}) = \prod_{k=0}^{n-1} \omega(g^k,g)$ (see Sec.~\ref{calcindex}) so (\ref{indexcal}) means that $\prod_{k=0}^{n-1} \omega(g^k,g) = 2^{\nu(P)}$. Also, recall that a coboundary transformation $\omega(g,h) \rightarrow \omega(g,h) \alpha(g) \alpha (h)/\alpha(gh)$, has the effect of multiplying $\prod_{k=0}^{n-1} \omega(g^k,g)$ by $\alpha(g)^{n}$. Since the fermionic GNVW index takes values in the set $\{\sqrt{2}^s p/q\}$\cite{fidkowski2019}, it follows that the most general coboundary transformation multiplies $\prod_{k=0}^{n-1} \omega(g^k,g)$ by a rational number of the form $\sqrt{2}^{ns} (p/q)^{n}$. If $2\nu(P)$ is not divisible by $n$, then $2^{\nu(P)}$ is not of this form. It then follows that $[\omega]$ is a nontrivial cocycle which means that $U_g$ is not onsiteable.

To complete the story, we now derive (\ref{indexcal}). The first step is to use the relationship between the GNVW index of a 1D locality preserving Gaussian unitary $U$ and the ``flow'' $\mathcal{F}(u)$ of the single particle unitary matrix $u_{jk}$ associated with $U$:
\begin{align}
    \mathrm{Ind}(U) = 2^{\mathcal{F}(u)}.
\label{flowid}
\end{align}
Here the matrix $u_{jk}$ is defined by $U^\dagger c_j U = \sum_{jk} u_{jk} c_k$, while the ``flow'' $\mathcal{F}(u)$, which takes values in $\mathbb{Z}$, is defined by choosing two adjacent but disjoint intervals $A,B$ that are large with respect to the range of $u$, and then defining $\mathcal{F}(u) = \sum_{a \in A, b \in B} (|u_{ab}|^2 - |u_{ba}|^2)$ \cite{kitaev2006}. The identity (\ref{flowid}) can be derived by considering the two special cases where $U$ is a unit translation or a finite depth Gaussian circuit: in the first case, the left and right hand sides both evaluate to $2$, while in the second case, both sides evaluate to $1$\cite{kitaev2006, fidkowski2019}. It then follows that the two sides must agree for any strictly locality preserving Gaussian unitary, since any such unitary is a composition of translations and finite depth Gaussian circuits\cite{fidkowski2019}.\footnote{We expect that a similar argument applies in the case where the unitary $U$ is not \emph{strictly} locality preserving and has superpolynomially decaying tails.} Substituting $U = U_{g,A}^{n}$ into (\ref{flowid}) gives 
\begin{align}
   \mathrm{Ind}(U_{g,A}^{n}) = 2^{\mathcal{F}(e^{2\pi i P_A})} .
\end{align}
At the same time, the results of Refs.~\onlinecite{rudner2013,zhanglevin2023} on the bulk-boundary correspondence for single particle unitary loops, imply that $\mathcal{F}(e^{2\pi i P_A}) = \nu(P)$. Putting these results together proves the claim (\ref{indexcal}).

\subsection{$\mathbb{Z}_2\times\mathbb{Z}_2^f$ SPT entangler}

Our second example is an anomaly-free, non-onsiteable $\mathbb{Z}_4$ fermionic symmetry with the interesting feature that it is also a (non-symmetric) entangler for the ``root" $\mathbb{Z}_2\times\mathbb{Z}_2^f$ SPT phase\cite{ryu2012,qi2013,yao2013,gu2014}. 

\begin{figure}[t]
    \centering
    \includegraphics[width=0.5\linewidth]{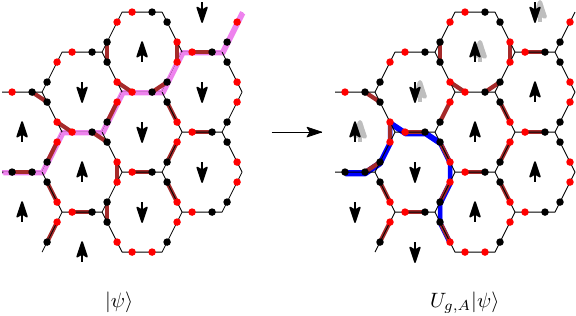}
    \caption{Truncated unitary $U_{g,A}$, where $A$ is the region below the violet line. From $|\psi\rangle$ to $U_{g,A}|\psi\rangle$, the (black) spins $\{\sigma_p^z\}$ on faces within $A$ are flipped; domain walls are identified as if spins on faces outside $A$ all point up ( grey arrows representing the up spins in $\{\tilde{\sigma}_p^z\}$ for $p\notin A$); Majorana fermions along the domain walls (blue lines) are translated by one, and if one Majorana is translated, the other one in the same pair is also translated. $U_{g,A}$ given this way preserves the total fermion parity.}
    \label{fig:fermion_restriction}
\end{figure}

We consider a Fisher lattice with one Majorana fermion on each vertex, and a qubit $\sigma_d$ on each  dodecagon face \cite{tarantino2016discrete}. This is equivalent to a honeycomb lattice with two Majorana fermions on each edge, and a qubit on each face, as shown in Figure \ref{fig:fermion_restriction}. On the original Fisher lattice, there is also a fictitious qubit $\sigma_t$ on each triangle $t$, its $\sigma_t^z$ value is $1$ or $-1$ depending on whether the majority of the three dodecagon faces bordering $t$ have $\sigma_d^z=1$ or $\sigma_d^z=-1$. There is a Kasteleyn orientation on all edges on the lattice, denoted by $\{s_{vw}\}$, where $s_{vw}$ take values in $1$ and $-1$, $v$ and $w$ are two vertices sharing an edge. The graph given by the Fisher lattice can also have dimer coverings. We denote any dimer covering by $\mathcal{D}=\{d_{vw}\}$, where $d_{vw}=1$ $(0)$ if the edge is (not) covered by a dimer. For each dimer covering of the edges, we can define a Hamiltonian such that the Majoranas connected by each dimer $d_{vw}=1$ are paired: $H_{\mathcal{D}} = -\sum_{\langle vw\rangle }i d_{vw}s_{vw}\gamma_v\gamma_w$. The ground states of the Hamiltonians $H_{\mathcal{D}}$ for all possible dimer covering $\mathcal{D}$ on this oriented Fisher lattice share the same total fermion parity.\cite{tarantino2016discrete}

We find that the following unitary defines a $\mathbb{Z}_4$ symmetry  that cannot be made on-site:
\begin{align}
U_g=\hat{S}[\{\sigma^z_f\}]\prod_d \sigma_d^x,
\end{align}
where $g$ is the generator of $\mathbb{Z}_4$, $d$ runs over all dodecagons, $f$ runs over both dodecagons and triangles, and $\hat{S}$ is the locality preserving unitary implementing a translation on the Majoranas on vertices based on the domain wall configuration of the plaquette qubits (including both dodecagon and triangle qubits). In particular, along plaquette domain walls, $\hat{S}$ implements a translation such that the forward direction has an $\uparrow$ domain to the left and a $\downarrow$ domain to the right, each Majorana is translated to its neighboring Majorana in the forward direction with a sign convention $\gamma_i\rightarrow s_{i,i+1}\gamma_{i+1}$, where $s_{i,i+1}$ is $1$ ($-1$) if the arrow from $i$ to $i+1$ (anti-)aligns with the Kasteleyn orientation on the edge connecting $i$ and $i+1$. The unitary maps the ground state of $H_{\mathcal{D}}$ to the ground state of $H_{\mathcal{D'}}$ with a different dimer covering, while preserving the total fermion parity. 

To see that $U_g$ generates a $\mathbb{Z}_4$ symmetry, note that $U_g^2$ maps every Majorana fermion $\gamma_i$ on each domain wall into $s_{i,i+1}s_{i+1,i}\gamma_i =-\gamma_i$, where the $(i+1)^\text{th}$ site is the neighbouring site of the $i^\text{th}$ site in the forward direction, and acts trivially on other Majorana fermions and all qubits $\sigma_d$. It follows that $U_g^4=1$.

We can choose the following way to truncate the unitary $U_g$, similar to our bosonic $\mathbb{Z}_2$ example, while preserving the total fermion parity symmetry. As we illustrate in Figure \ref{fig:fermion_restriction}, for a region $A$, to implement $U_{g,A}$, we first flip the spins on all dodecagons, and then identify the domain walls for $\{\tilde{\sigma}_d^z\}$, where $\tilde{\sigma}_d^z=\sigma_d^z$ for $d\in A$, and $1$ otherwise. We perform a translation along the domain wall in such a way that if one Majorana is translated along the domain wall, the Majorana that is paired up with it is also translated along the domain wall. In this way, after the action of $U_{g,A}$, the Majorana fermions remain paired according to a dimer covering on the Fisher lattice, thus leaving the total fermion parity unchanged. 

One can further compute $\Theta=\operatorname{Ind} (U^4_{g,A})$ shown in Eq.~(\ref{eq:Theta}) and Eq.~(\ref{eq:Theta_derivation}) and find that $\Theta=2$, which is not a fourth power of a rational number or $\sqrt{2}$, or their combinations. Thus, the $\mathbb{Z}_4$ symmetry carries an index given by a non-trivial element in $H^2(\mathbb{Z}_4, \mathbb{Q}_+[\sqrt{2}])$, where $\mathbb{Q}_+[\sqrt{2}]=\left\{q\cdot {\sqrt{2}}^{n}|q\in \mathbb{Q}_+,n\in \mathbb{Z}\right\}.$ 
We believe that $U_g$ is equivalent to a bosonic non-onsitable $\mathbb{Z}_4$ symmetry up to an FDQC, with possible tensoring with suitable ancilla degrees of freedom.

Meanwhile, if one applies $U_g$ to a trivial product state $|\psi_0\rangle = \otimes_{e=\langle vw\rangle} |i s_{vw}\gamma_v\gamma_w=1\rangle \otimes_{d}|\sigma_d^x=1\rangle$, where $e$ labels edges of the honeycomb lattice, then in the output state, any domain wall defined by $\{\sigma_d^z\}$ is decorated by a one-dimensional state in the Kitaev phase \cite{kitaev2001unpaired}. This gives the ground state of the ``root'' $\mathbb{Z}_2\times \mathbb{Z}_2^f$ SPT state \cite{tarantino2016discrete}, where the $\mathbb{Z}_2$ symmetry $\prod_d\sigma^x_d$ flips the spins $\sigma_d^z$. Nevertheless, the entangler $U_g$ does not commute with the $\mathbb{Z}_2$ symmetry generated by $\prod_{d}\sigma_d^x$. To see this, one can for example consider the action of $U_g \prod_{d}\left(\sigma_d^x\right) U_g^{-1}$ on a single Majorana fermion operator $\gamma_i$ on a $\mathbb{Z}_2$ domain wall by conjugation, and finds that $\gamma_i$ becomes $s_{i,i+1}s_{i+1,i+2}\gamma_{i+2}$, rather than remaining invariant. 


\bibliography{onsite_draft.bib}